\documentclass[12pt,preprint]{aastex}
\usepackage{graphicx}
\begin{document}
\def\la{\mathrel{\hbox{\rlap{\hbox{\lower4pt\hbox{$\sim$}}}\hbox{$<$}}}}
\def\ga{\mathrel{\hbox{\rlap{\hbox{\lower4pt\hbox{$\sim$}}}\hbox{$>$}}}}
\def\lam{$\lambda$}
\def\kms{km~s$^{-1}$}
\def\vphot{$v_{\rm phot}$}
\def\ang{~\AA}
\def\syn{SYNOW}
\def\dm15{{$\Delta$}$m_{15}$}
\def\rsi{$R$(Si~II)}
\def\v10{$V_{10}$(Si~II)}
\def\wsi{$W_lambda$(Si~II)}
\def\vdot{\.v(Si~II)}
\def\W575{$W(5750)$}
\def\W610{$W(6100)$}
\def\6100{the 6100~\AA\ absorption}
\def\tex{$T_{\rm exc}$}
\def\ve{$v_{\rm e}$}

\title {Comparative Direct Analysis of Type~Ia Supernova Spectra.
IV. Postmaximum}

\author {David Branch\altaffilmark{1},
 David~J. Jeffery\altaffilmark{1,2}, Jerod Parrent\altaffilmark{1,3},
 E.~Baron\altaffilmark{1}, M.~A. Troxel\altaffilmark{1},
 V.~Stanishev\altaffilmark{4}, Melissa Keithley\altaffilmark{1},
 Joshua Harrison\altaffilmark{1}, and Christopher
 Bruner\altaffilmark{1,5}}

\altaffiltext{1} {Homer L. Dodge Department of Physics and Astronomy,
University of Oklahoma, Norman,~OK; branch@nhn.ou.edu}

\altaffiltext{2} {Department of Physics, University of Idaho,
Moscow,~ID}

\altaffiltext{3} {Department of Physics and Astronomy, Dartmouth
College, Hanover,~NH}

\altaffiltext{4} {Department of Physics, Stockholm University,
Stockholm, Sweden}

\altaffiltext{5} {Department of Physics, Worcester Polytechnic
Institute, Worcester,~MA}

\begin{abstract}

A comparative study of optical spectra of Type~Ia supernovae (SNe~Ia)
obtained near 1 week, 3 weeks, and 3 months after maximum light is
presented.  Most members of the four groups that were defined on the
basis of maximum light spectra in Paper~II (core normal, broad line,
cool, and shallow silicon) develop highly homogeneous postmaximum
spectra, although there are interesting exceptions.  Comparisons with
SYNOW synthetic spectra show that most of the spectral features can be
accounted for in a plausible way.  The fits show that 3 months after
maximum light, when SN~Ia spectra are often said to be in the nebular
phase and to consist of forbidden emission lines, the spectra actually
remain dominated by resonance scattering features of permitted lines,
primarily those of Fe~II.  Even in SN~1991bg, which is said to have
made a very early transition to the nebular phase, there is no need to
appeal to forbidden lines at 3 weeks postmaximum, and at 3 months
postmaximum the only clear identification of a forbidden line is
[Ca~II] \lam\lam7291, 7324.  Recent studies of SN~Ia rates indicate
that most of the SNe~Ia that have ever occurred have been ``prompt''
SNe~Ia, produced by young ($\sim 10^8$ yr) stellar populations, while
most of the SNe~Ia that occur at low redshift today are ``tardy'',
produced by an older (several Gyrs) population.  We suggest that the
shallow silicon SNe~Ia tend to be the prompt ones.

\end{abstract}


\section{INTRODUCTION}

This is the fourth in a series of papers on a comparative direct
analysis of the optical spectra of Type~Ia supernovae (SNe~Ia).
Paper~I (Branch et~al. 2005) was concerned with a time series of
spectra of the spectroscopically normal SN~1994D, Paper~II (Branch
et~al. 2006) was devoted to spectra obtained near maximum light, and
Paper~III (Branch et~al. 2007) concentrated on premaximum spectra.
This paper is about postmaximum spectra, which form in the deeper,
lower velocity ejecta.

In Paper~II we divided the maximum-light ($B$ maximum) spectra into
four groups: core normal, broad line, cool, and shallow silicon
(denoted CN, BL, CL, and SS, respectively, in this paper).  The group
assignments were made on the basis of measurements of the (pseudo)
equivalent widths of absorption features near 5750 and 6100\ang, as
well as on the appearance (depth, width, and shape) of the 6100\ang\
absorption, which is produced by the Si~II \lam6355 transition.
Although we framed the presentation and discussion in terms of the
four groups, in the end we concluded that for the most part the
spectra appeared to have a continuous distribution of properties,
rather than consisting of discrete subtypes (the extreme SS
SN~2002cx--likes being an apparent exception, and the extreme CL
SN~1991bg--likes being a possible exception).  In Paper~III we found
that to a large extent the premaximum spectra exhibited the defining
characteristics of the four groups.

In this paper, as in the previous ones, we confine our attention to
optical spectra, from the Ca~II H and K feature in the blue
($\sim3700$\ang) to the Ca~II infrared triplet (Ca~II IR3) in the red
($\sim9000$\ang).  All spectra have been corrected for the redshifts
of the host galaxies, and mild smoothing has been applied to some of
the spectra.  All spectra have been flattened according to the local
normalization procedure of Jeffery et~al. (2007), which eliminates all
significant broad band continuum variations, including those caused by
interstellar reddening and observational error.  Thus the locally
normalized spectra allow valid comparisons of intrinsic line features.
But note that a locally normalized spectrum is not a unique
representation of the line spectrum.  It depends on the exact
prescription of the local normalization procedure.  We use exactly the
same procedure for all spectra of this paper.

We examined all of the SN~Ia postmaximum spectra available to us and
selected three samples: a ``1 week postmax'' sample, consisting of one
spectrum each of 21 SNe~Ia observed between day~+6 and day~+8 with
respect to the time of maximum brightness in the $B$ band; a ``3 week
postmax'' sample of 19 SNe~Ia observed between day~+19 and day~+23;
and a ``3 months postmax'' sample of 15 SNe~Ia observed between
day~+81 and day~+98 (an interval of day~+80 to day~+100 yielded no
additional spectra).  The SNe~Ia and the epochs of the selected
spectra are listed in Table~1.  Many comparisons of subsets of these
spectra have appeared in the literature.  Our goal is to provide a
more systematic and comprehensive comparison.

Spectra of four of the SNe~Ia of Table~1 were not available for
previous papers of this series.  Sadakane et~al. (1996) published
spectra of SN~1995D obtained shortly after maximum light, but it is
difficult to decide from their Figure~3 whether SN~1995D should be
assigned to the CNs or the SSs.  On the basis of the report by Benetti
\& Mendes de Oliveira (1995) that in a 1 week premax spectrum the
5750~\AA\ absorption was unusually weak if present at all, we
tentatively assign SN~1995D to the SS group.  Pastorello et~al. (2007)
referred to SN~2004eo as a transitional SN~Ia because its properties
placed it between the three groups (faint, low velocity gradient, and
high velocity gradient) of Benetti et~al. (2005).  In our
classification, the maximum light spectrum of SN~2004eo is not quite
CN; it resembles that of SN~1989B (Paper~II), and like SN~1989B it is
placed in the CL group although it does not contain the blue ``Ti~II
trough'' of the more extreme members of the CL group.  The maximum
light spectrum of SN~2006X (Wang et~al. 2007) is that of an extreme BL
SN~Ia, such as SN~1984A.  Hicken et~al. (2007) showed that an early
spectrum of SN~2006gz had the strongest C~II \lam6580\ absorption yet
seen in a SN~Ia, and argued that super--Chandrasekhar mass ejection is
required.  Although no spectrum was obtained within three days of
maximum light, at earlier epochs SN~2006gz showed characteristics of a
SS event, and we tentatively classify it as such.

\section{CALCULATIONS}

Continuing our attempt to provide an internally consistent
quantification of SN~Ia spectra, we have used the parameterized
resonance--scattering synthetic--spectrum code, SYNOW, to fit the
spectra of Table~1.  The excitation temperature is fixed at a nominal
value of 7000~K, as was done for the postmaximum SN~1994D spectra of
Paper~I. A change of procedure from our previous work with SYNOW on
postmaximum spectra, including that of Paper~I, is that instead of
using exponential (Paper~I) or power--law radial optical depth
distributions, we use a flat distribution with an imposed maximum
velocity, $v_{\rm max}$, for every ion.  In the outer layers of the
ejected matter, a decreasing density is dictated by hydrodynamical
models (and by mass conservation: the density must start decreasing
somewhere), but in the deeper layers the density gradient is less
steep.  Since the optical depth distribution depends also on
composition, excitation, and ionization gradients, a flat optical
depth distribution is not unreasonable.  For further economy of
parameters, for a given spectrum we use the same value of $v_{\rm
max}$ for all singly ionized members of the iron group, from Sc~II to
Ni~II (with the exception of the 1 week postmax spectrum of SN~2000cx;
\S~3.)  The price to be paid is that the fits are not optimized, but
the fitting procedure becomes more efficient (there is no need to vary
the $v_{\rm e}$ parameter) and the parameters become easier to
compare: for each ion we have just a reference--line optical depth and
the velocity interval in which the lines of the ion are forming in the
synthetic spectrum, the minimum velocity being either the velocity at
the photosphere $v_{\rm phot}$ or a higher detachment velocity (i.e.,
detached from the photosphere).  Since we use only one optical depth
component for each ion, in this paper it is not necessary to use the
HV (high velocity) and PV (photospheric velocity) terminology of
previous papers of this series.

Consider the effects, when switching from an exponential to a flat
optical depth distribution, on the relative importance of lines of the
same ion but of different strengths.  In the flat case, the optical
depths of the strongest lines need to be smaller than in the
exponential case, since they maintain the same optical depths all the
way to $v_{\rm max}$.  Lines that are somewhat weaker, but whose
optical depths still remain above unity in the flat case, become
relatively more important, because they now form over just as large a
velocity interval as the strongest lines.  Lines whose optical depths
are reduced from above unity in the exponential case to well below
unity in the flat case no longer conspicuously affect the spectrum.

As an example, Figure~1 compares synthetic spectra for Fe~II, with
flat and exponential optical depth distributions.  The flat case has
the parameters that we use in \S~4 for the day~+22 spectrum of the CN
SN~1996X: $\tau$(Fe~II)=12, $v_{\rm phot} = 6000$ \kms, and $v_{\rm
max}=13,000$ \kms.  The exponential case has the parameters that we
used in Paper~I for the day~+24 spectrum of the CN SN~1994D:
$\tau$(Fe~II)=200, $v_{\rm phot} = 9000$ \kms, and exponential
$e$--folding velocity $v_e = 1000$ \kms.  Note that the flat case
requires a lower value of $v_{\rm phot}$ than the exponential case.
While the two synthetic spectra are similar, there are some
significant differences, such as the near disappearance in the flat
case of the absorption that appears near 5380~\AA\ in the exponential
case, and the higher flux peak near 4660~\AA\ in the flat case.

\section{ONE WEEK POSTMAX}

One week postmax is in the middle of what we referred to in Paper~I as
the postmaximum ``Si~II phase'' (from two to 12 days past maximum, for
SN~1994D), because the 6100~\AA\ absorption is deep and apparently
unblended, at least in its core.  At 1 week postmax the spectra are
not radically different from at maximum light.

Figure~2 shows the 1 week postmax spectra of the CNs.  The spectra are
very similar.  A log plot such as Figure~2 is convenient for comparing
multiple spectra at once, but since the spectra are displaced from
each other, it can be difficult to appreciate the degree of the
homogeneity.  Figure~3 directly compares the spectra of SN~1998bu and
SN~1996X, to show how strikingly similar these two spectra are.  Even
the difference at the red end of the SN~1998bu spectrum is not
necessarily real, because observed spectra sometimes have problems at
their ends, and also because the local normalization technique can
mildly warp the ends of the spectra (Jeffery et~al. 2007).

An example SYNOW fit is shown in Figure~4.  The fit is to the spectrum
of the CN SN~1996X (selected because it is the best spectrum of
Figure~2, not because it is the best fit).  The fitting parameters for
SN~1996X (and other selected spectra of Figure~2) are in Table~2.
Apart from the flat optical depth distribution, the fit of Figure~4 is
a conventional one.  The ions are the same as were used for the 1 week
postmax spectrum of SN~1994D in Paper~I, except that here Cr~II also
is included.  Lines of O~I, Si~II, S~II, Ca~II, and Fe~II produce most
of the features in the synthetic spectrum, and we are confident that
these ions are present in the observed spectrum.  The presence in the
observed spectrum of Na~I, Mg~II, Cr~II, Fe~III, and Co~II is not
definite, but we use them in the synthetic spectrum because they are
plausible and they improve the fit.

The three main discrepancies in Figure~4 are familiar problems with
SYNOW fits. First, for Ca~II we choose parameters to fit the observed
IR3 feature, because it is less blended and more sensitive to optical
depth than the H and K feature; in this case, this causes the
synthetic H and K absorption to be too strong.  Second, the flux
minimum from about 6900 to 7100~\AA\ has no counterpart in the
synthetic spectrum.  The only identification that we can suggest is
[O~II] \lam\lam7320,7330, which was discussed in the context of
SN~1991T by Fisher et~al. (1999) and of the Type~Ic SN~1994I by
Millard et~al. (1999).  However, at later epochs this discrepancy
becomes stronger, and invoking [O~II] to solve it would imply
excessive mass and kinetic energy of oxygen (at least for an exploding
white dwarf).  More likely, what we are seeing in Figure~4 is an
early, mild manifestation of a discrepancy that is due to our
simplifying assumption of resonant scattering.  Third, the synthetic
spectrum lacks absorption from about 7600 to 7800~\AA.  The observed
absorption in this region presumably is due to Mg~II \lam7890, blended
with O~I \lam7772, but in SYNOW spectra the synthetic absorption
usually is too blue, even when Mg~II is undetached, as it is here.

Most of the spectra of Figure~2 are so similar that it is not worth
reporting fitting parameters for every spectrum.  The exception is
SN~2004S, which has much stronger high velocity Ca~II absorption than
the others.  Krisciunas et~al. (2007) termed SN~2004S a clone of
SN~2001el, another CN with exceptionally strong high velocity Ca~II
features (Wang et~al. 2003; Kasen et~al. 2003; Mattila et~al. 2005;
Paper~II).  Krisciunas et~al. noted that at maximum light SN~2004S had
an unusually low Si~II absorption blueshift for a spectrum containing
the usual SN~Ia features, and that the blueshift decreased more
rapidly than in SN~2001el.  We find that at 1 week postmax, all
absorptions other than those of Ca~II are less blueshifted than in the
other spectra of Figure~2; consequently our fit for SN~2004S has a low
value of $v_{\rm phot}= 7000$ \kms\ (Table~2), instead of 11,000 \kms\ as
used for the others of Figure~2.

Figure~5 shows the 1 week postmax spectra of the BLs (plus the
spectrum of the CN SN~1996X for comparison).  At this epoch the
spectrum of SN~2002er is like that of a CN.  SN~1992A retains some of
its BL characteristics and remains mildly different from CN.  The
spectra of SN~2002bf, SN~2006X, and SN~1984A are obviously different
from CN, but similar to each other.  Their 6100~\AA\ absorptions are
deeper, broader, and more blueshifted, and they have deep absorptions
from about 4700 to 5100~\AA.  A fit to SN~2002bf is shown in Figure~6.
The 6100~\AA\ absorption and the deep absorption from 4700 to
5100~\AA\ are matched by using a high $v_{\rm max}$ of 21,000 \kms\
for Si~II and Fe~II, compared to a typical value of 15,000 \kms\ for
the CNs at 1 week postmax.  The Fe~II features of SN~2005bf are so
prominent that it is difficult to determine whether lines of Mg~II,
Fe~III, and Co~II are present.

The 1 week postmax spectra of the CLs are shown in Figure~7.  At this
epoch SN~1989B is like that of a CN and SN~2004eo is only mildly
different.  The spectrum of the extreme CL SN~1999by, a
SN~1991bg--like (Garnavich et~al. 2004), is more highly evolved than
the others of the 1 week postmax sample, in fact it has more in common
with the 3 week postmax spectra of the CNs than with their 1 week
postmax spectra.  A fit for SN~1999by is shown in Figure~8.  The
presence of Si~II, Ca~II, Ti~II, and Fe~II is definite, while Mg~I,
Sc~II, and Cr~II are plausible and improve the fit.  There are three
main discrepancies.  First, the synthetic absorption near 5340~\AA,
produced by Cr~II, is not deep enough, even though Cr~II is too strong
in several other places.  (A corresponding feature in the CNs of the 3
week postmax sample is discussed in \S~4.)  Second, the observed
absorption near 5820~\AA\ has practically no counterpart in the
synthetic spectrum.  The only identification we can offer is Si~II
\lam5972, but it hardly appears in the synthetic spectrum even though
the absorption feature of the \lam 6355 Si~II reference line is too
strong.  Third, the synthetic flux peaks near 4000 and 4600~\AA\ are
too high; this may be a consequence of the flat optical depth
distribution (see the 4600~\AA\ peak in Figure~1.)  Our present SYNOW
fit differs from the one presented in Garnavich et~al. in that we do
not use Ca~I, Ti~II makes no significant contribution to the synthetic
spectrum at wavelengths longer than 5300~\AA, and we do invoke Mg~I,
Sc~II, and Cr~II.

Figure~9 shows the 1 week postmax spectra of the SSs.  SN~2006gz is
intermediate between the CN SN~1996X and the SS SN~1999aa, which has
weaker Si~II, S~II, and Ca~II absorptions.  Hicken et~al. (2007)
identified C~II in spectra of SN~2006gz obtained 10 or more days
before maximum light, after which the feature was not identifiable,
consistent with the lack of evidence for C~II in the spectrum shown in
Figure~9.  The direct comparison of SN~1999ac and SN~1996X in
Figure~10 is interesting.  SN~1999ac has Si~II and S~II absorptions
that are less blueshifted than in SN~1996X, but Fe~II features from
4700 to 5100~\AA\ that are more blueshifted.  A fit for SN~1999ac is
shown in Figure~11.  The presence of O~I, Si~II, S~II, Ca~II, and
Fe~II in SN~1999ac is definite; Na~I, Mg~II and Fe~III are plausible.
To account for the above mentioned peculiarity, we use a low
photospheric velocity of 6000 \kms\ and a higher $v_{\rm max}$ for
Fe~II (15,000 \kms) than for most of the other ions (see Table~2).
The spectra of SN~1999ac have been discussed extensively by Garavini
et~al. (2005).  They emphasized the unusually low Si~II blueshift,
consistent with our use of a low photospheric velocity.  They also
concluded that Fe~II lines formed at unusually low velocities, but our
$v_{\rm max}$ of 15,000 \kms\ is the same as we use for the CN
SN~1996X.

The Ca~II IR3 feature of SN~2000cx obviously differs from the others
of Figure~9.  Otherwise, the closest match to SN~2000cx is SN~1999aa,
but there are conspicuous differences.  The fit for SN~2000cx shown in
Figure~12 is unusual because, as in Paper~II, we resort to detached
high--velocity Ti~II, Cr~II, and Fe~II.  In fact, in the synthetic
spectrum most of the photospheric features are at least mildly
detached (Table~2).  Still, the fit is not very good.  Although the
synthetic Ca~II H and K absorption is much too strong, the IR3 does
not appear.  The observed multicomponent absorption of the IR3 has
been discussed by Li et~al. (2001), Wang et~al. (2003), Kasen
et~al. (2003), and Thomas et~al. (2004), and is known to have involved
asymmetrical structures at high velocity.

\section{THREE WEEKS POSTMAX}

Three weeks postmax is in the middle of what we referred to in Paper~I
as the ``Si~II--to--Fe~II transition phase'' (from 14 to 28 days past
maximum, for SN~1994D) because the core of the 6100~\AA\ absorption is
present but flanked by strengthening Fe~II features.  The 3 week
postmax spectra are very different from the 1 week postmax spectra.

Figure~13 shows the 3 week postmax spectra of the CNs, and a fit for
SN~1996X is shown in Figure~14.  The ions are the same as were used
for SN~1994D at comparable epochs in Paper~I.  Apart from the Ca~II
and Na~I features, the dominant ion is Fe~II, but Si~II, Cr~II, and to
a lesser extent Co~II are also needed.  Fitting parameters for
SN~1996X and other selected spectra of the 3 week postmax sample are
given in Table~3.

The degree of homogeneity in Figure~13 is very high, except in two
respects.  First, SN~2001el and SN~2004S still have strong high
velocity Ca~II H and K absorptions (and most of the absorption
features of SN~2004S continue to be somewhat less blueshifted than in
the others).  Second, the relative heights of the flux peaks that
flank the 5350~\AA\ absorption are not all the same.  The relative
heights of these two peaks were discussed in Paper~I: in SN~1994D the
ratio of the peak on the right of the 5350~\AA\ absorption to the peak
on the left increased steadily from day~+15, when the peaks were
roughly equally high, to day~+28, when the peak on the right was much
higher.  Our fits accounted nicely for this evolution, in terms of
strengthening Fe~II and Cr~II lines.  In this respect, some of the
spectra of Figure~13 appear to be spectroscopically ``earlier'' than
some of the others.  For example, the day~+21 spectrum of SN~1998aq
(Fig.~13) has equally high peaks, like the day~+15 spectrum of
SN~1994D (Paper~I), and closely resembles that spectrum in other
respects also.  Even CN spectra can get mildly out of phase, with
respect to the time of B--band maximum.

The 3 week postmax spectra of the BLs are shown in Figure~15.  At this
epoch SN~1981B is like CN and so is SN~2002er, apart from its deep
high velocity Ca~II H and K absorption (which is not matched by the
appearance of the IR3 absorption).  Even the extreme BL SN~1984A
appears to differ only mildly from CN, although the limited wavelength
coverage does not allow us to see how the Ca~II lines are behaving.

Figure~16 shows the 3 week postmax spectra of the CLs.  At this epoch
SN~1989B and SN~2004eo are like CN.  SN~1986G remains mildly
different, e.g., its 5750~\AA\ absorption is shallow, and the ratio of
the flux peaks flanking the 5350~\AA\ absorption is very large.
SN~1991bg continues to show obvious differences.  A fit for SN~1991bg
is shown in Figure~17.  All 7 ions used in the synthetic spectrum are
considered to be definite.  In the blue the synthetic spectrum is a
complex blend of Ti~II, Cr~II, and Fe~II.  Not only does SN~1991bg
have Ti~II while CNs do not, but also Cr~II plays a more important
role than it does in the CNs.

The 3 week postmax spectra of the SSs are shown in Figure~18.
SN~1999ee is similar to CN, although with a strong high velocity Ca~II
H and K absorption.  SN~1999aa also is similar to CN except for the
spectroscopically ``early'' ratio of the flux peaks flanking the
5350~\AA\ absorption.  In this respect, as in others, the day~+19
spectrum of SN~1999aa in Figure~18 is a good match to the day~+14
spectrum of SN~1994D (Paper~I).

As shown in previous papers (Chornock et~al. 2006; Phillips
et~al. 2007; Stanishev et~al. 2007), at multiple epochs the spectra of
SN~2002cx and SN~2005hk are very similar.  In Figure~18, the spectra
of both have been artifically blueshifted by 5000 \kms, to
approximately align their absorption features with those of the other
SSs.  The direct comparison in Figure~19 of SN~2005hk and the CN
SN~1996X shows that although the features of SN~2005hk are narrower,
when the spectrum of SN~2005hk is artificially blueshifted, the
features of the two spectra show a strong correspondence.  Not only
the absorptions but also the flux peaks match up well.  If the flux
peaks of SN~2005hk were true emission peaks, at the rest wavelengths
of the lines that produce them, then after the blueshifting, the peaks
would be bluer than those of SN~1996X.  This is in accord with the
maxim that in heavily blended resonance--scattering spectra,
``absorptions trump emissions'', i.e., flux peaks do not necessarily
occur at the rest wavelengths of the strongest lines, because they are
strongly influenced by absorption components of mainly redward lines.
Figure~20 shows a fit to the (not artificially blueshifted) spectrum
of SN~2005hk.  The ions used are the same ones as used for the 3 week
postmax spectrum of the CN SN~1996X, except that Si~II is not needed
for SN~2005hk.  Note that it is not $v_{\rm phot}$, but $v_{\rm max}$
for Cr~II, Fe~II, and Co~II that differs by 5000 \kms\ from that of
SN~1996X (Table~2).  Apart from the two Ca~II features and the one
Na~I feature, the synthetic spectrum is a complex blend of Fe~II,
Cr~II, and Co~II, just as it is for SN~1996X.

As can be seen in Figure~18, at this epoch SN~2000cx remains different
from CN in several ways, e.g., the peaks that appear near 4800 and
4920~\AA\ in SN~1996X are smeared out in SN~2000cx.  In this respect,
SN~2000cx resembles the BL SN~2002bf of the 1 week premax sample
(\S~2).  To fit the spectrum of SN~2000cx we use a high maximum
velocity of 18,000 \kms\ for Fe~II and Cr~II (Table~2). A discussion
of SYNOW fits to day +15 and +32 spectra, which have better wavelength
coverage, can be found in Branch et~al. (2004).

\section{THREE MONTHS POSTMAX}

Three months postmax is within what we referred to in Paper~I as the
``Fe~II phase'' (from 50 to 115 days after maximum for SN~1994D, with
these particular limiting epochs being determined by the availability
of spectra) because the 6100~\AA\ absorption is nearly gone at 50 days
and the spectrum is mainly shaped by Fe~II lines.  The 3 months
postmax spectra are quite different from the 3 weeks postmax spectra.
The spectra of the five CNs (Fig.~21) are very much alike.  It is
often assumed that at this epoch the spectrum is composed of optically
thin, collisionally excited forbidden lines, but the SYNOW fit for
SN~2003du in Figure~22 is good enough to firmly establish that
whatever the underlying source of the emission, permitted lines of
Na~I, Ca~II, and Fe~II are mainly responsible for shaping the spectrum
(see also the discussion in Paper~I).  However, our fit is very poor
from 6600 to 7800~\AA.  It may be that the broad flux minimum near
6750~\AA\ is not an absorption feature, and that the flux peaks from
7270 to 7700 are produced by strong net emission.  Bowers
et~al. (1997) modeled the day~+95 optical spectrum (along with the
day~92 infrared spectrum) of SN~1995D assuming that the spectrum
consists only of forbidden--line emissions.  Their identifications of
the flux peaks are shown in Figure~22 (SN~1995D and SN~2003du have the
same peaks, and for some of the peaks Bowers et~al. did not suggest
identifications).  In the blue, the Bowers et~al. fit was not
successful.  However, our poor fit from 6600 to 7800~\AA\ may signal
that in this wavelength range forbidden emission lines are emerging
without strong modification by permitted--line scattering.  On the
other hand, if we make the Fe~II lines stronger, thereby making our
fit worse in the blue, we do begin to obtain a better fit in the 6600
to 7800~\AA\ interval.  This could be a consequence of a wavelength
dependent variation in the depth of the photosphere.  At present, line
identifications in this wavelength range are uncertain.

At this epoch the four BL events of the 3 months postmax sample
(Fig.~23), including the extreme BL SN~2006X, are similar to CN.

Among the CLs (Fig.~24), SN~1989B is like CN, SN~1986G differs mildly,
and the extreme CL SN~1991bg remains distinctly different.  The fit
for SN~1991bg in Figure~25 shows that permitted--line scattering,
mainly by Fe~II, shapes the spectrum in the blue, but the fit fails
from 6000 to 8000~\AA.  The only clear evidence for forbidden lines is
the strong emission near 7250~\AA, due to [Ca~II] (Filippenko
et~al. 1992a).

At this epoch the probable SS SN~1995D and the extreme SS SN~1991T
(Fig.~26) are like CN, and even the maverick SN~2000cx is only mildly
different.

\section{DISCUSSION}

The spectra that appear in this series of papers are simply the good
spectra that are available to us, so the samples are affected not only
by observational bias in favor of supernovae that are bright and have
slowly declining light curves, but also by the observer bias in favor
of obtaining and publishing spectra of unusual events.  Nevertheless,
these samples clearly indicate a trend toward increasing spectroscopic
conformity at later epochs.  Among the 24 SNe~Ia of the maximum light
sample (Paper~II), only 7 were admitted to the (by construction)
highly homogeneous CN group.  By contrast, of the 12 SNe~Ia of the 3
months postmax sample, only SN~1991bg is distinctly different from the
CNs.  (SN~2002cx--likes would be different too, if they were in the
sample.)  This spectroscopic convergence suggests that although the
outer layers of SNe~Ia are diverse in various ways, the deep layers of
most SNe~Ia are fundamentally the same.

In this paper, the line identifications for the CNs are much like they
were in Paper~I for SN~1994D, although the fitting parameters are
different because of our present use of flat optical depth
distributions.  The conclusion that permitted lines strongly affect
the spectrum as late as 3 months postmax leads to a caveat to the
statement above that the deep layers of most SNe~Ia are fundamentally
the same: at three months postmax the strongest permitted lines are
strong, which entails line profile saturation and a limited
sensitivity of the spectra to physical differences in the ejecta.  If
the spectra of the 3 months postmax sample consisted simply of
optically thin forbidden emission lines, saturation would not be an
issue and the observed spectroscopic homogeneity would imply a
startling (unbelievable?) degree of physical homogeneity.  The degree
of diversity should increase at still later times as the permitted
lines weaken and the forbidden lines emerge unscathed.  The transition
from resonance--scattering to forbidden--line dominated spectra will
be the subject of a future paper.

As in previous papers of this series, we find that the BL SNe~Ia have
essentially the same line identifications as the CNs, but with more
line optical depth at high velocities.  The distinct characteristics
of the extreme BL events SN~1984A, SN~2002bf, and SN~2006X at 1 week
postmax raises the issue of whether they are a discrete subgroup, but
only in terms of the properties of their outer layers.  There is no
evidence that their deep layers are unusual: by 3 months postmax,
SN~2006X was like a CN.

The spectrum of the extreme CL SN~1991bg has been thought to have
undergone an early transition to the nebular phase (Filippenko
et~al. 1992a; Leibundgut et~al. 1993; Turatto et~al. 1996; Mazzali
et~al. 1997).  According to our interpretation (Fig.~17) there is no
need to appeal to forbidden lines in the 3 week postmax spectrum
(Fig. 17), and even in the 3 month postmax spectrum the only obvious
one is [Ca~II] \lam\lam7291, 7324.  In spite of its peculiarities, the
spectrum of SN~1991bg does increasingly resemble the spectra of other
SNe~Ia after maximum light, and in all respects SN~1986G appears to be
a link between SN~1991bg--likes and normal SNe~Ia.  As discussed in
Paper~I, the spectroscopic peculiarities of SN~1991bg--likes may
largely (but not entirely) reflect a temperature threshold below which
key ionization ratios change abruptly.  The issue of whether
SN~1991bg--likes are a distinct subgroup remains open.

The 1 week and 3 week postmax spectra of the SS SN~2000cx are unique
among SNe~Ia observed so far.  Our identifications of high velocity
Ti~II and especially Cr~II are not definite, but for nonstandard
spectra, nonstandard identifications are to be expected, and no
alternative identifications have been suggested.  The resemblance of
the 3 months postmax spectrum of SN~2000cx to CNs suggests, though,
that its deeper layers are like those of other SNe~Ia.  The
SN~2002cx-likes, on the other hand, appear to be radically different
from other SNe~Ia, even in their deeper layers.  Pending the discovery
of SNe~Ia having less mild SN~2002cx--like properties, the
SN~2002cx--likes appear to be a discrete subgroup.  Yet the
resemblance of their spectra, when artificially blueshifted, to other
SNe~Ia (Fig.~19) suggests kinship.\footnote{An anonymous referee
points out that another possible discrete subgroup could be SNe~Ia
that have very slow light curve decline rates, such as SN~2001ay
(Howell \& Nugent 2004), SN~2003fg (Howell et~al. 2006), and SN~2006gz
(Hicken et~al. 2007).  Subclassifying SNe~Ia by means of photometric
characteristics is beyond the scope of this paper, but we note that in
our spectroscopic classification scheme, SN~2001ay is broad line
(Paper~II) while SN~2003fg (Paper~III) and SN~2006gz (this paper) are
shallow silicon.}

Although this series of papers is confined to spectroscopy, it is
worth emphasizing that the spectroscopic diversity is bound to
contribute to the photometric diversity.  Branch et~al. (2004)
discussed the effects of the putative high velocity Ti~II of SN~2000cx
on the evolution its $B-V$ color.  Another good example is the 1 week
postmax spectra of the extreme BLs SN~2002bf, SN~1984A, and SN~2006X
(Figures 5 and 6).  The deep Fe~II absorption from 4700 to 5100~\AA,
near the interesection of the B and V bands, throws flux right into
the peak of the V band (thus obscuring the S~II absorptions) and is
bound to affect the B-V color evolution, with implications for
attempts to estimate the interstellar extinction on the basis of broad
band photometry (e.g., Wang et~al. 2007 for SN~2006X).

In recent years it has been shown (Mannucci et~al. 2005, 2006;
Scannapieco \& Bildsten 2005; Sullivan et~al. 2006) that some SNe~Ia,
called prompt, are produced by a young ($\sim 10^8$ yr) stellar
population while others, called tardy, are produced by an older
population (several Gyr).  Most the SNe~Ia that have ever occurred
were prompt, but at the present epoch most of the SNe~Ia at low
redshift are tardy.  This raises the question of which of our
spectroscopic groups are prompt and which are tardy.  Since the
extreme CL events such as SN~1991bg tend to occur in old populations
and the extreme SS events such as SN~1991T tend to occur in young
populations (Sullivan et~al. 2006 and references therein), if we are
to break our groups into a minority of prompts and a majority of
tardies, then the members of the SS group should tend to be prompts
and the members of the other three groups tend to be tardies. Yet
there is no obvious spectroscopic separation between the SS group and
the CN group.  The CN group was defined to emphasize the strong
spectroscopic homogeneity of its members, but several of the SSs
(e.g., SN~1999ee, SN~1999ac, SN~1999aa) are not very different from
CN.  As discussed in Paper~II, it may be that there are two
evolutionary paths to SNe~Ia that require different amounts of time
but produce two families of SNe~Ia that have overlapping distributions
of their properties.

That being said, Quimby et~al. (2007) suggested that even among normal
SNe~Ia (in the broad sense of the term) there may be two distinct
classes of events: those that have a smoothly declining density
distribution in their outer layers and therefore show a blueshift of
the 6100~\AA\ absorption that decreases smoothly with time, and those
in which silicon is largely confined to a shell--like density
structure and therefore have a more constant blueshift of the
6100~\AA\ absorption.  The smooth density structure is characteristic
of deflagration and delayed detonation models, while a shell is
characteristic of pulsating delayed detonation and tamped detonation
models.  The suggestion of Quimby et~al. was motivated by a prolonged
period of nearly constant blueshift in SN~2005hj, but they point out
that some other SNe~Ia such as SN~1999aa and SN~2000cx are similar in
this respect.  In our classification, these three events are SS.  More
well observed SS and CN SNe~Ia are needed to determine whether SS and
CN SNe~Ia are distinctly different in this respect.

SYNOW is useful for establishing line identifications and information
on the velocity intervals in which the lines are forming, but it is no
substitute for detailed calculations of spectra of hydrodynamical
models.  At present no 1D explosion model has been to shown to account
well for the observed spectral evolution, not even for the CNs (Baron
et~al. 2006).  Spectropolarimetry and explosion modeling tells us that
asymmetries are present, but the spectra of 3D explosion models cannot
yet even be calculated in full detail (although a start has been made
by calculating the spectra of 3D models for angle averaged
compositions; Baron et~al. 2007).  The homogeneity of the CNs shows
that there is a standard, repeatable SN~Ia explosion model that does
not have large compositional inhomogeneities near or below the maximum
photospheric velocity of about 12,000 \kms\ --- but what is this
model?  And there are many additional questions, such as: what
variation on the CN model is responsible for the extra high velocity
matter in the BLs, and the putative bizarre high velocity features of
SN~2000cx, and the distinctive properties of the SN~2002cx--likes?  We
have hardly begun on the path toward understanding the various
manifestions of spectroscopic diversity of SNe~Ia.

We are grateful to all observers who have provided spectra.  This work
has been supported by NSF grants AST 0506028 and AST 0707704, NASA
LTSA grant NNG04GD36G, and DOE grant DEFG02-07ER41517.

\clearpage     

\begin{figure}
\includegraphics[width=.8\textwidth,angle=270]{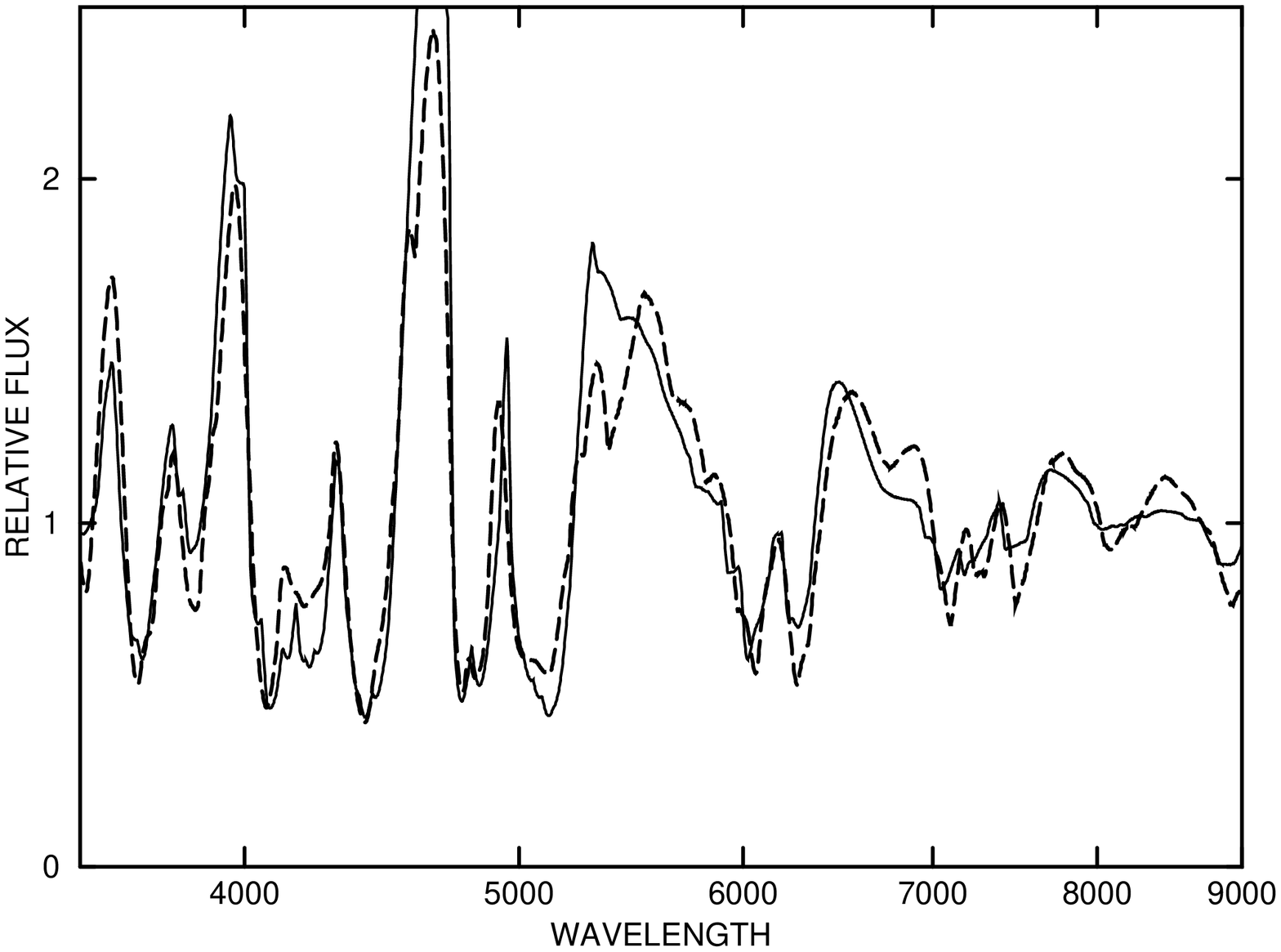}
\caption{Synthetic spectra for Fe~II with flat ({\sl solid line}) and
  exponential ({\sl dashed line}) optical depth distributions are
  compared.  The parameters for the flat case are $\tau$(Fe~II)=12,
  $v_{\rm phot} = 6000$ \kms, and $v_{\rm max}=13,000$ \kms.  For the
  exponential case they are $\tau$(Fe~II)=200, $v_{\rm phot} = 9000$ \kms,
  and $v_e = 1000$ \kms.}
\end{figure}

\begin{figure}
\includegraphics[width=.8\textwidth,angle=0]{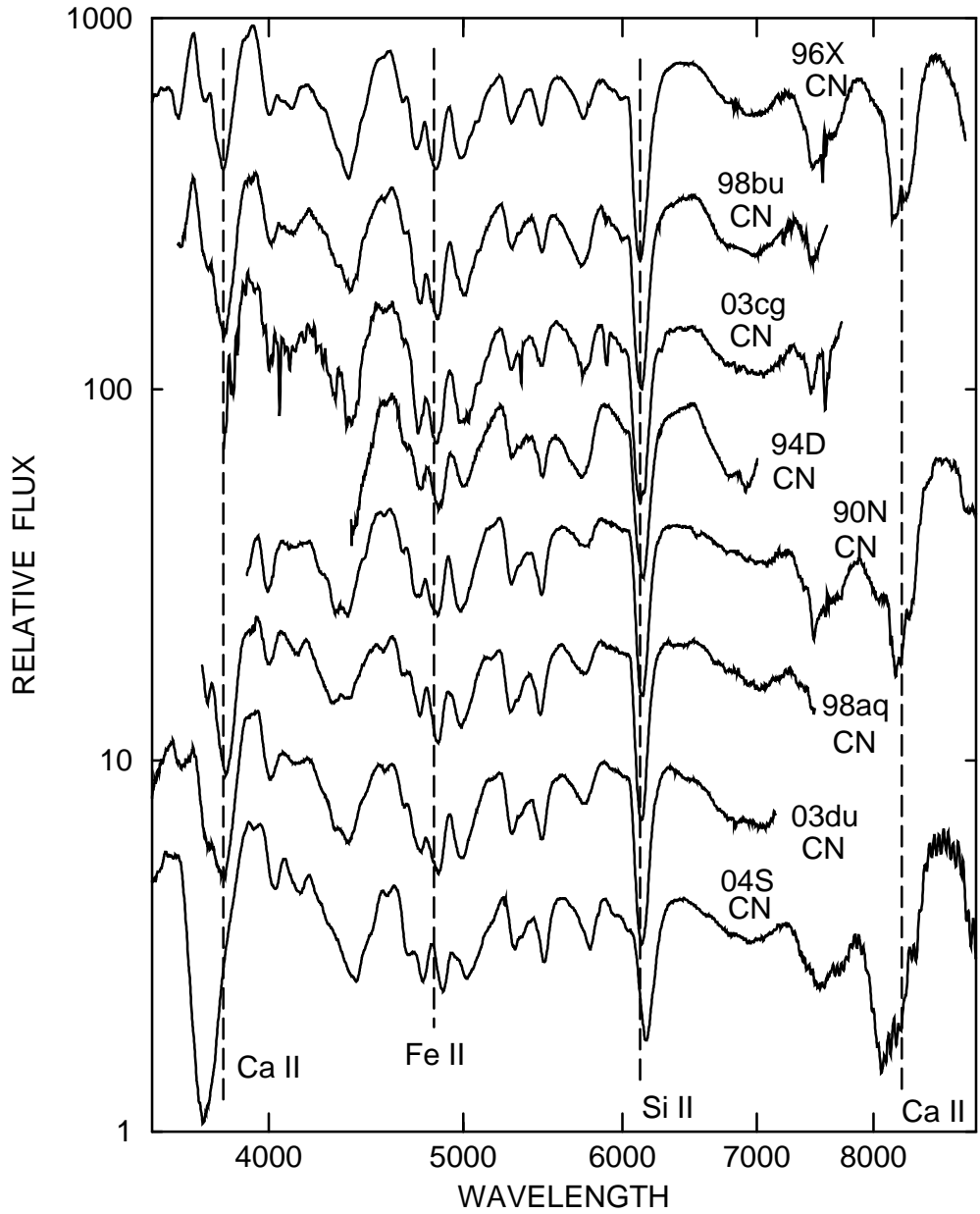}
\caption{Spectra of the eight CNs of the 1 week postmax
sample.  The spectra have been flattened according to the local
normalization technique of Jeffery et~al. (2007).  Vertical
displacements are arbitrary and narrow absorptions near 7600\ang\ and
6900\ang\ are telluric.  Vertical dashed lines refer to Ca~II
\lam3945, Fe~II \lam5018, Si~II \lam6355, and Ca~II \lam8579,
blueshifted by 11,000 \kms.}
\end{figure}

\begin{figure}
\includegraphics[width=.8\textwidth,angle=270]{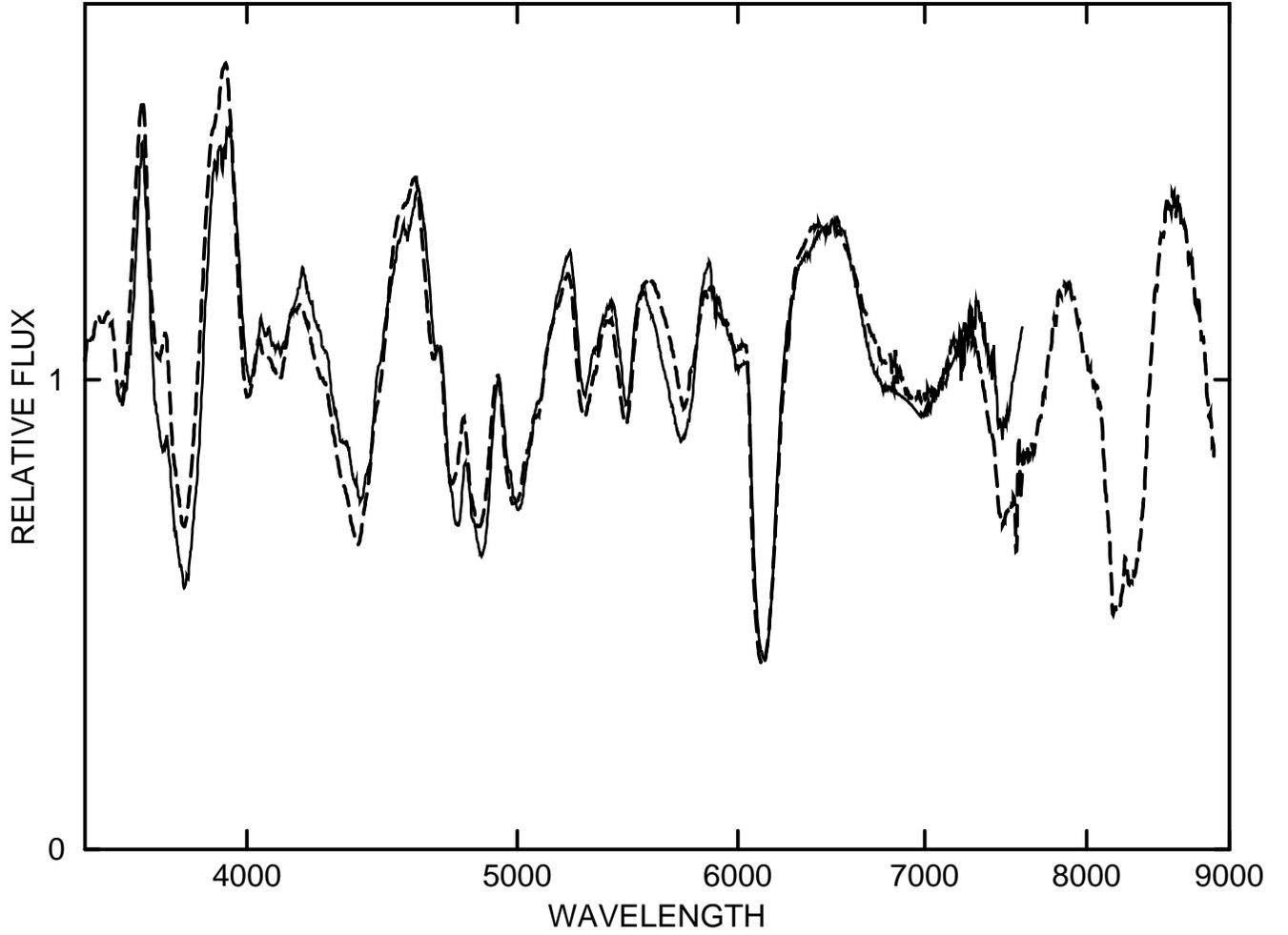}
\caption{ The 1 week--postmax spectra of the CN SN~1998bu ({\sl solid
  line}) from Jha et~al. (1999) and the CN SN~1996X ({\sl dashed
  line}) from Salvo et~al. (2001) are compared.}
\end{figure}

\begin{figure}
\includegraphics[width=.8\textwidth,angle=270]{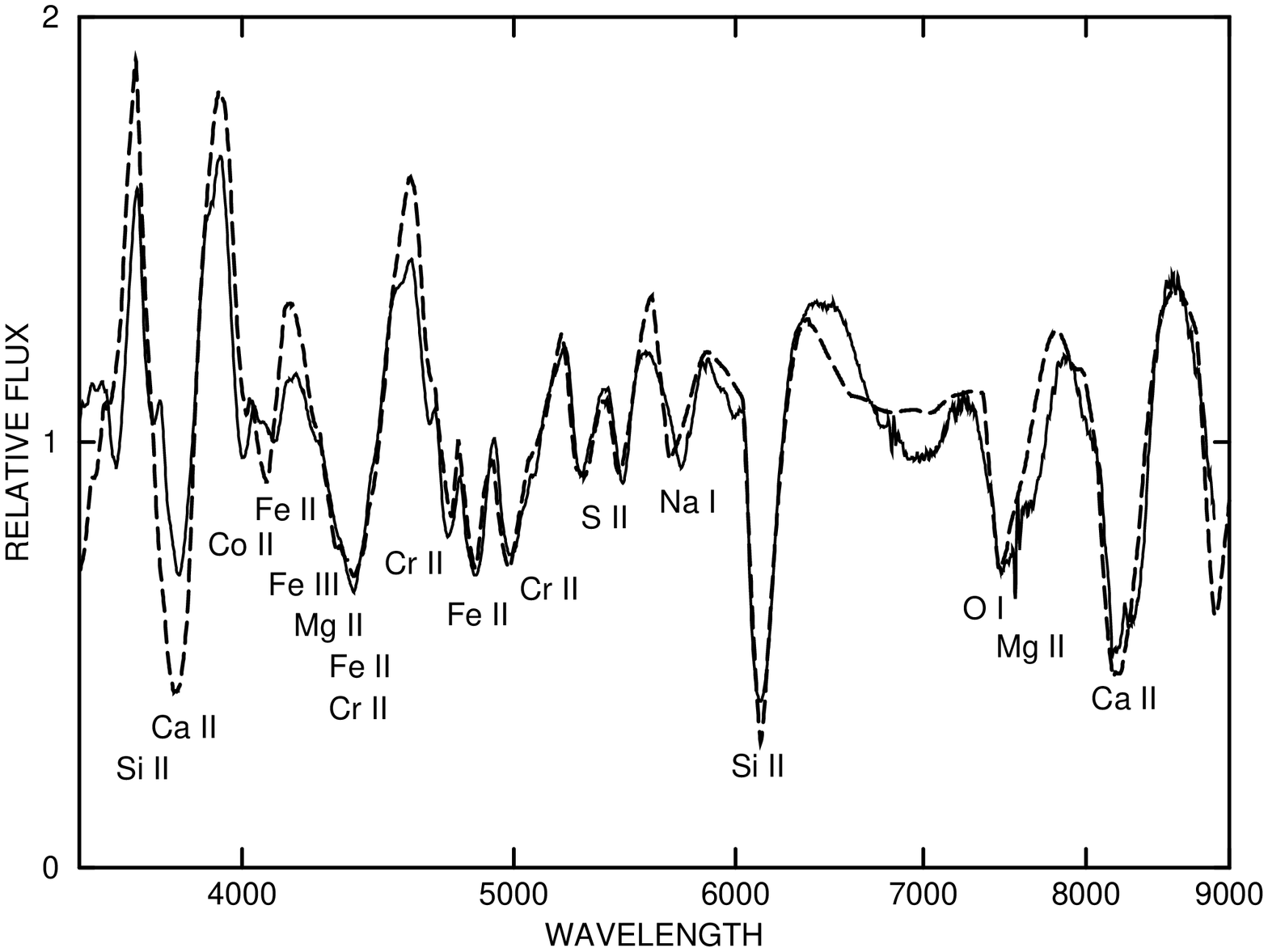}
\caption{The 1 week postmax spectrum of the CN SN~1996X ({\sl
  solid line}) from Salvo et~al. (2001) compared with a synthetic
  spectrum ({\sl dashed line}).}
\end{figure}


\begin{figure}
\includegraphics[width=.8\textwidth,angle=0]{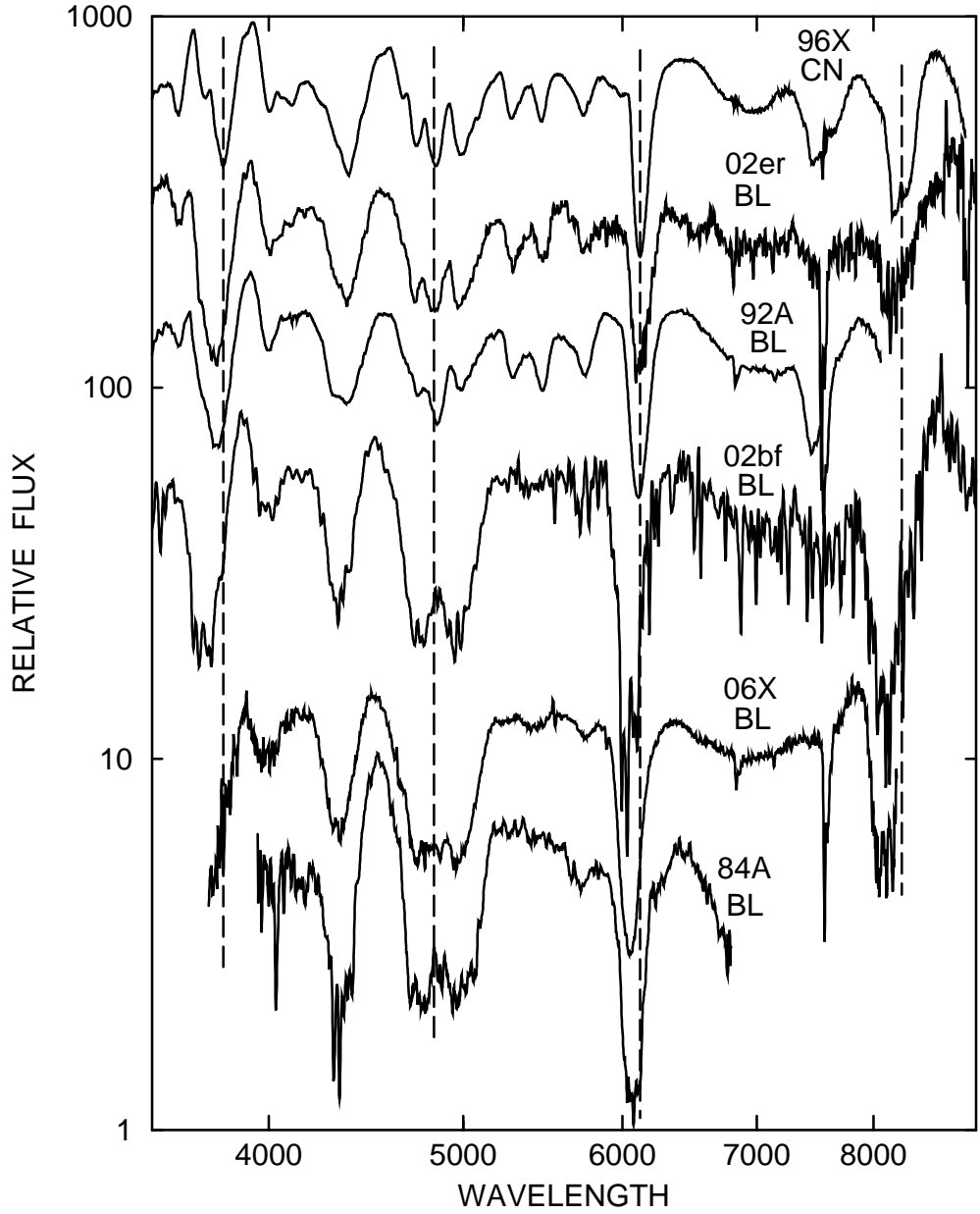}
\caption{Like Fig.~2 but for one CN (for comparison) and
the five BLs of the 1 week postmax sample.}
\end{figure}

\begin{figure}
\includegraphics[width=.8\textwidth,angle=270]{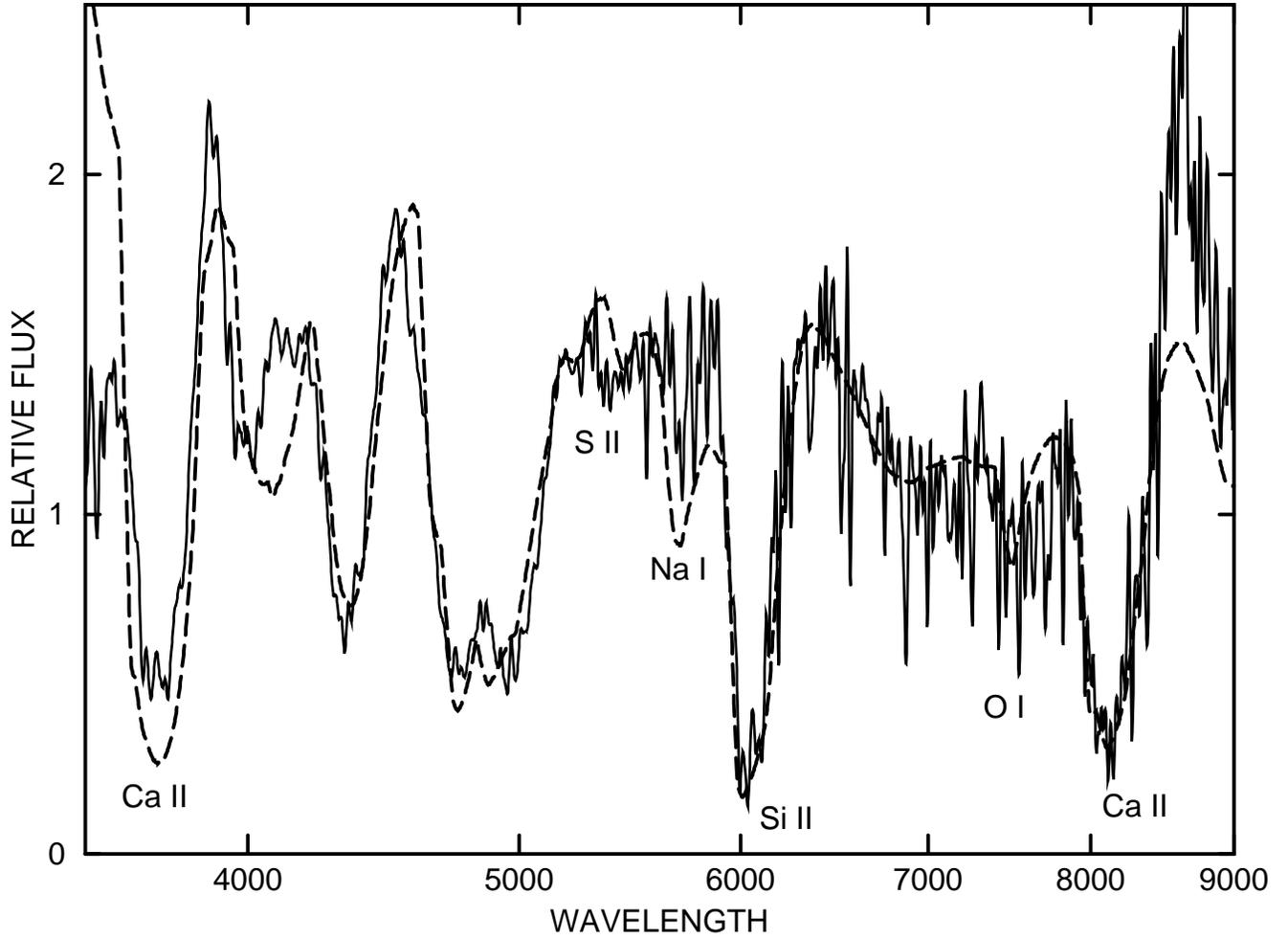}
\caption{The 1 week postmax spectrum of the BL SN~2002bf ({\sl solid
  line}) from Leonard et~al. (2005) compared with a synthetic spectrum
  ({\sl dashed line}).  Unlabelled absorption features in the
  synthetic spectrum are produced by Fe~II.}
\end{figure}

\begin{figure}
\includegraphics[width=.8\textwidth,angle=0]{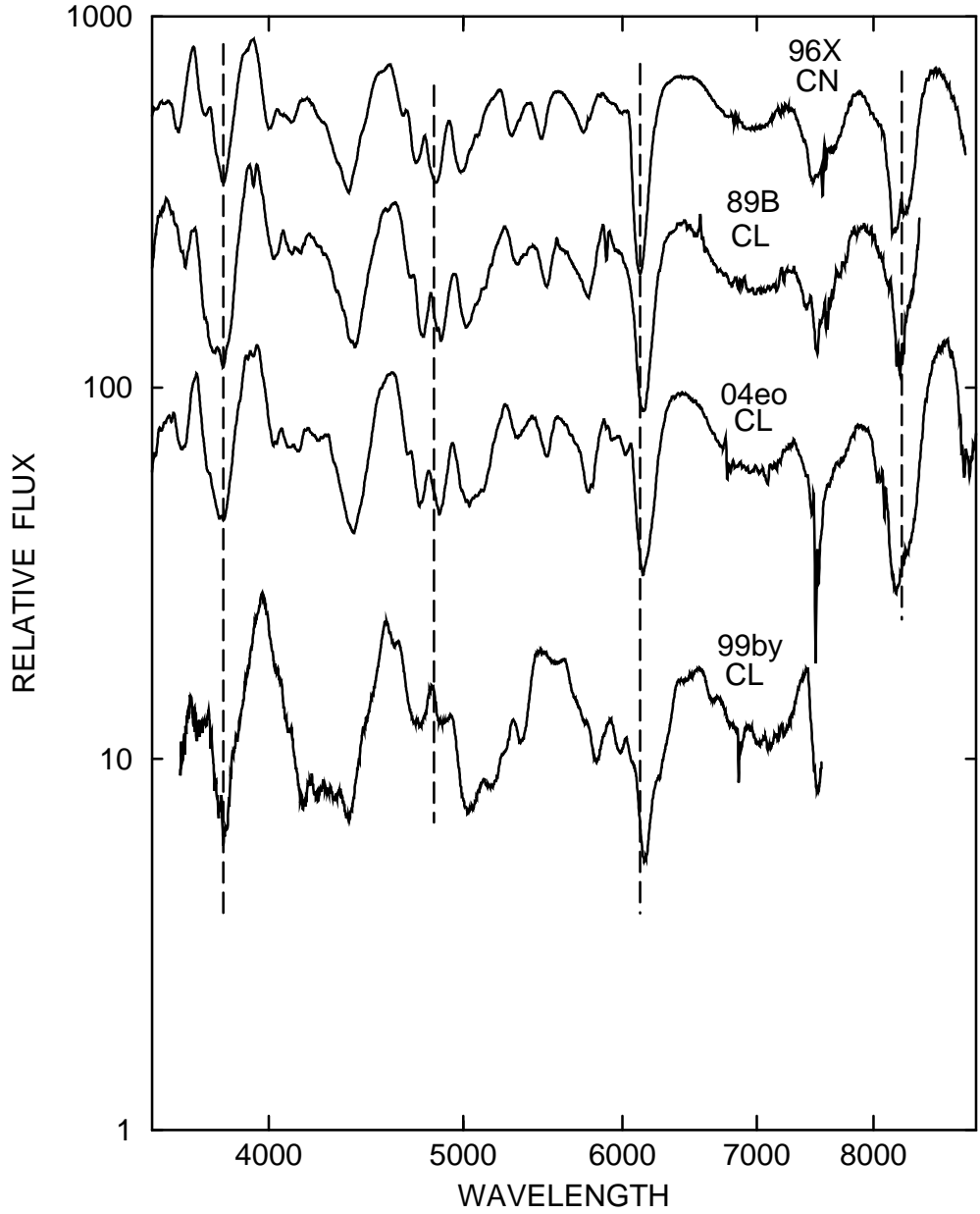}
\caption{Like Fig.~2 but for one CN and the three CLs of the
1 week postmax sample.}
\end{figure}

\begin{figure}
\includegraphics[width=.8\textwidth,angle=270]{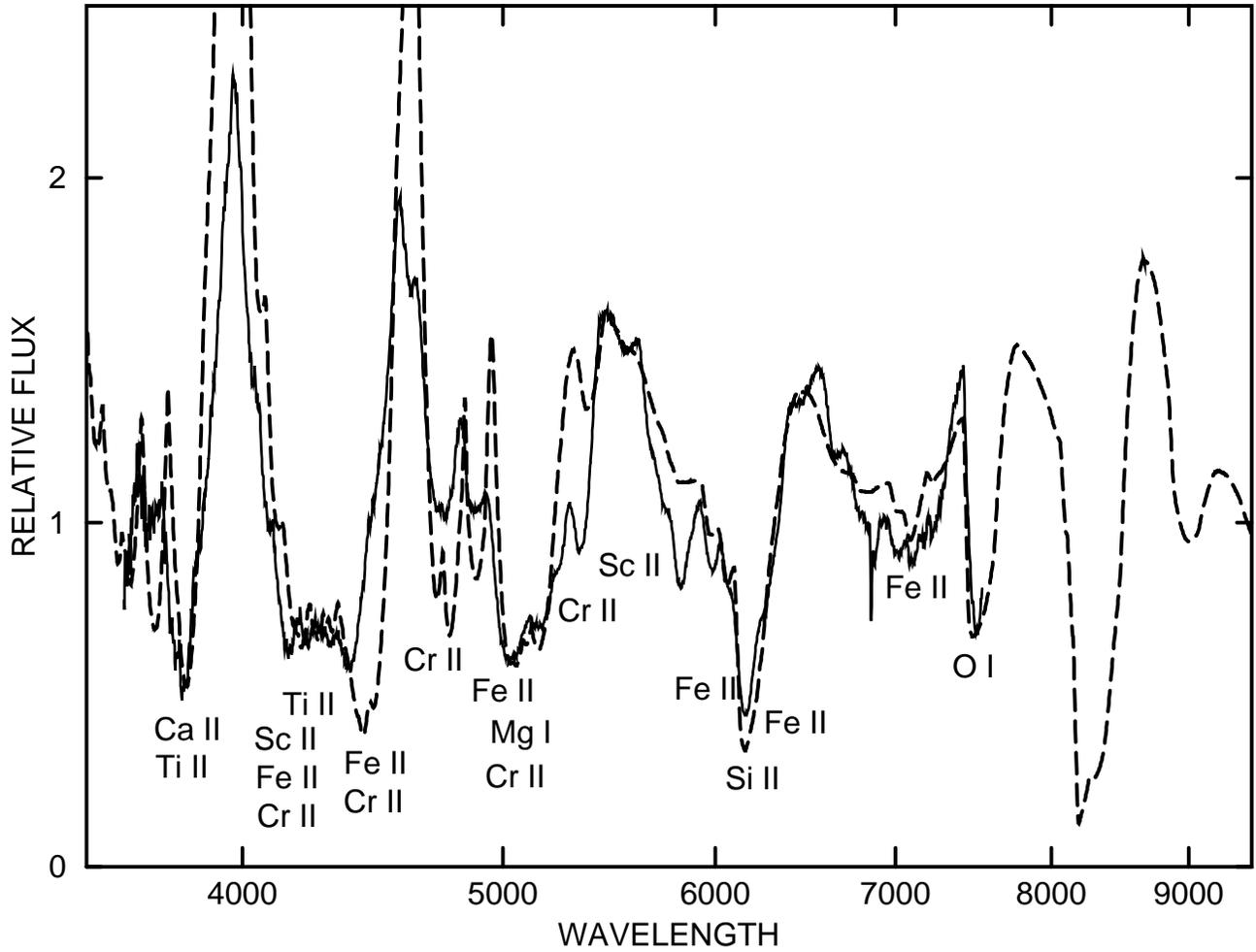}
\caption{The 1 week--postmax spectrum of the CL SN~1999by ({\sl
  solid line}) from Garnavich et~al. (2004) compared with a synthetic
  spectrum ({\sl dashed line}).}
\end{figure}

\begin{figure}
\includegraphics[width=.8\textwidth,angle=0]{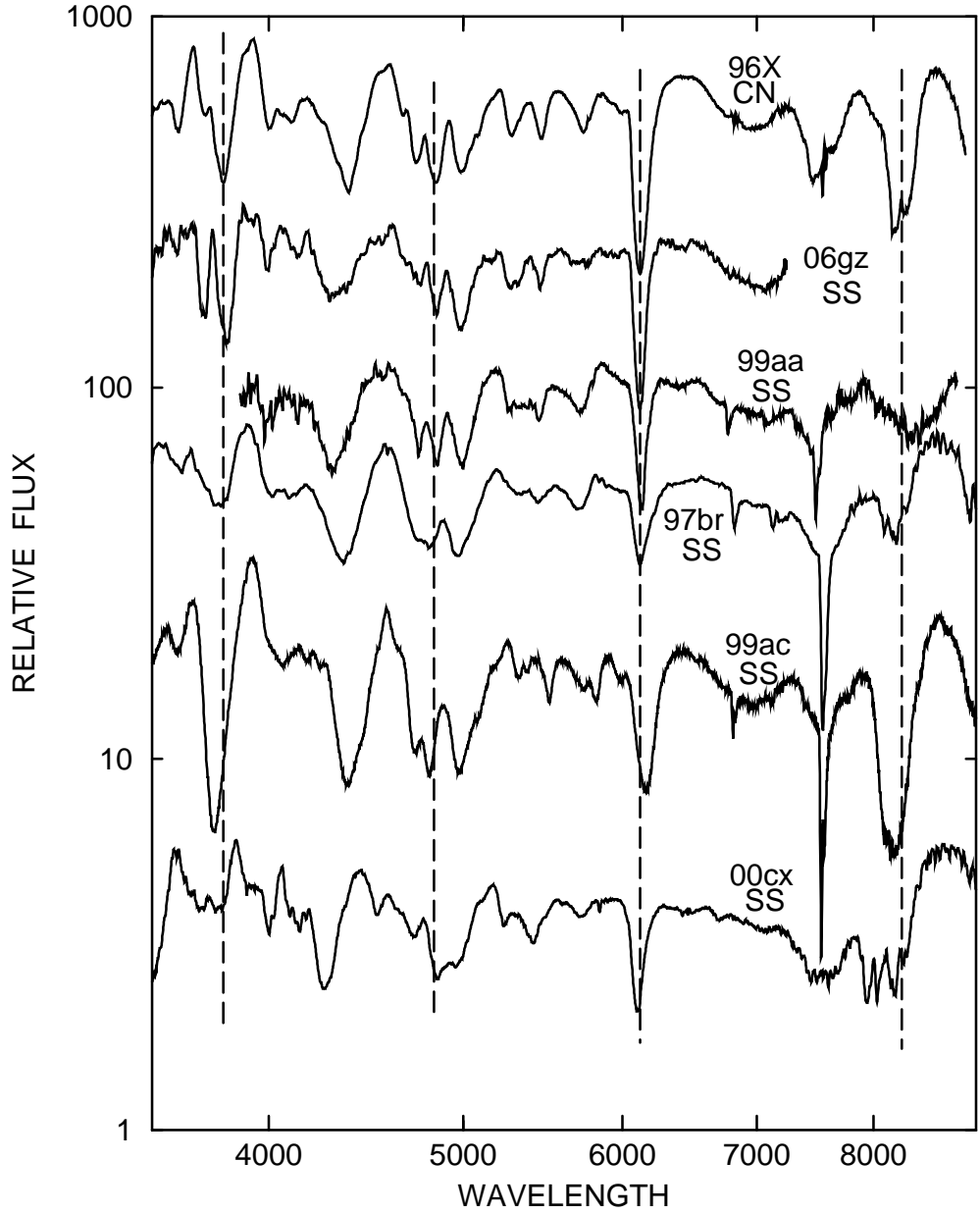}
\caption{Like Fig.~2 but for one CN and the five SSs of the 1 week
postmax sample.}
\end{figure}

\begin{figure}
\includegraphics[width=.8\textwidth,angle=270]{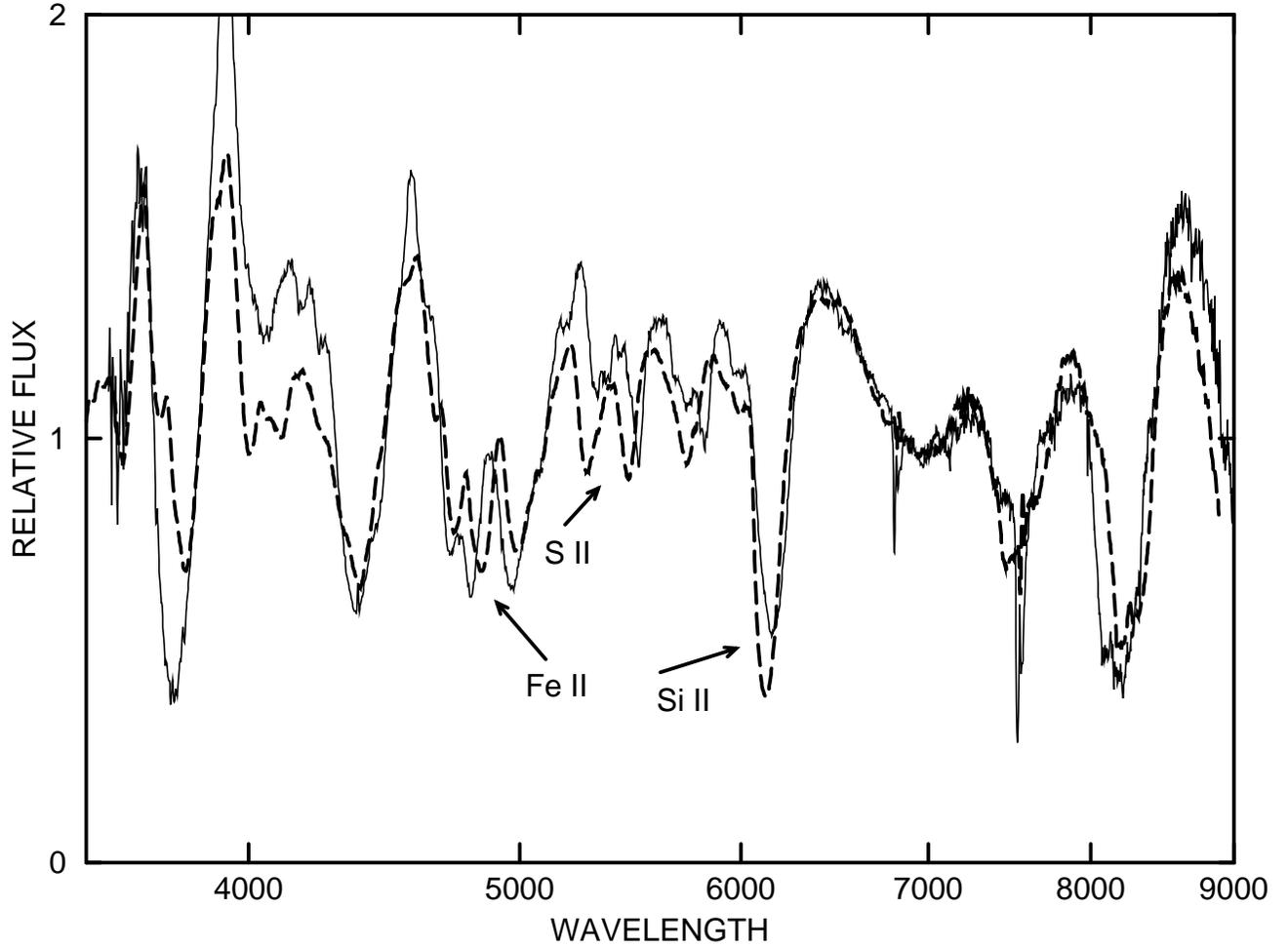}
\caption{ The 1 week postmax spectra of the SS SN~1999ac
  ({\sl solid line}) from Garavini et~al. (2005) and the CN
  SN~1996X ({\sl dashed line}) from Salvo et~al. (2001) are compared.
The arrows point to the lower blueshifts of Si~II and S~II and the
  higher blueshifts of Fe~II, in SN~1999ac.}
\end{figure}

\begin{figure}
\includegraphics[width=.8\textwidth,angle=270]{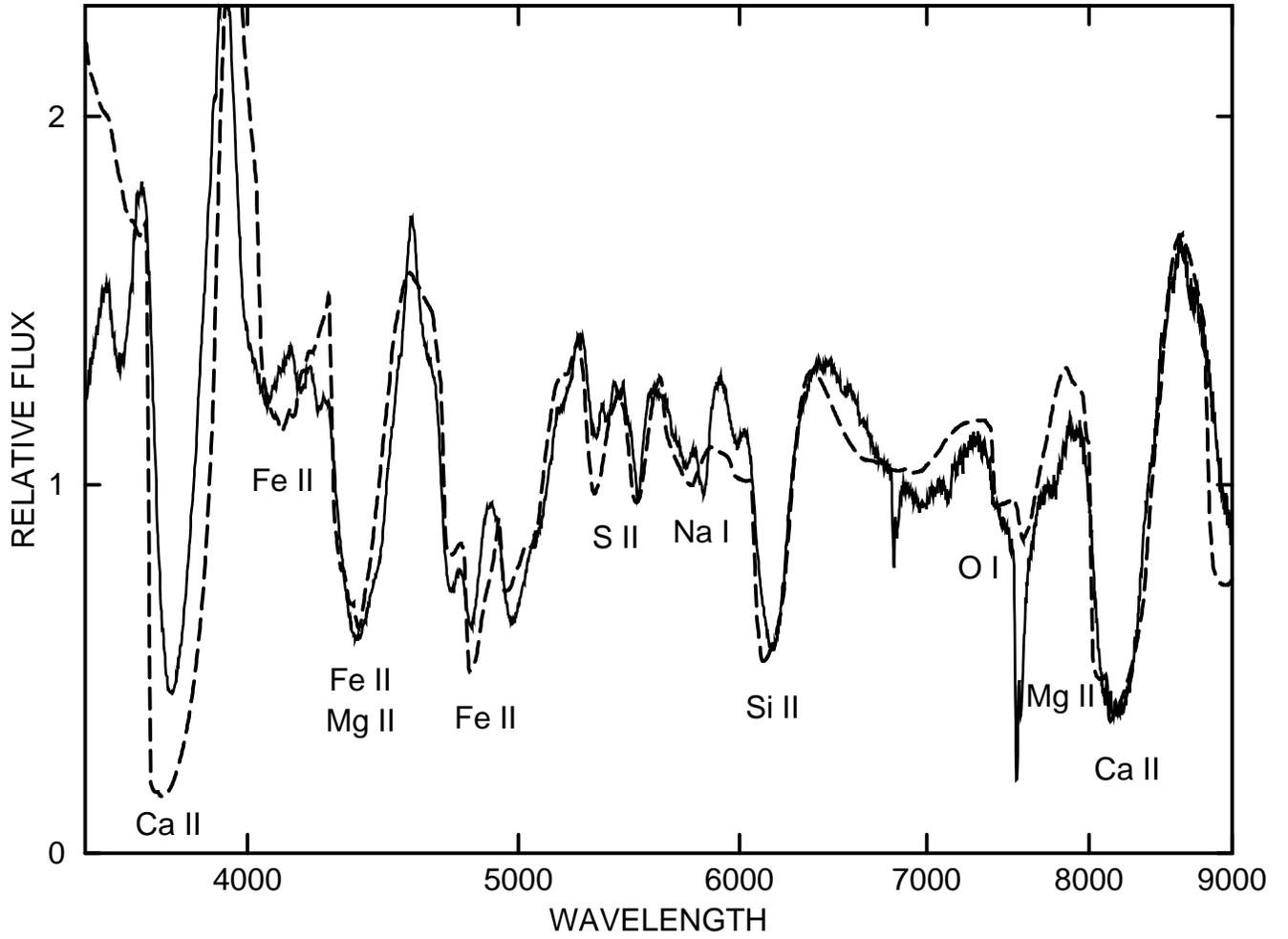}
\caption{The 1 week postmax spectrum of the SS SN~1999ac
  ({\sl solid line}) from Garavini et~al. (2005) compared with a
  synthetic spectrum ({\sl dashed line}).}
\end{figure}


\begin{figure}
\includegraphics[width=.8\textwidth,angle=270]{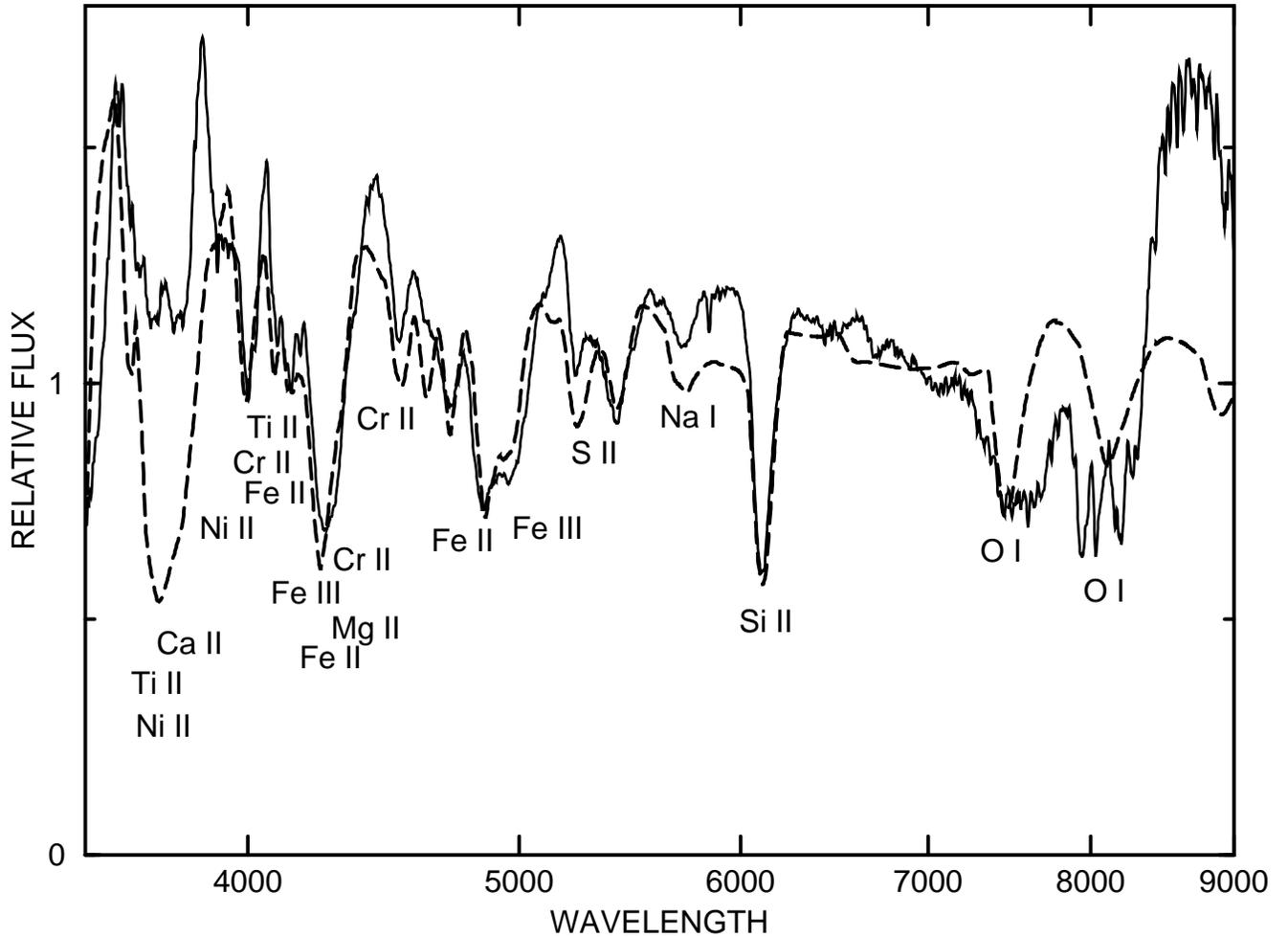}
\caption{The 1 week postmax spectrum of the SS SN~2000cx ({\sl solid
  line}) from Li et~al. (2001) compared with a synthetic spectrum
  ({\sl dashed line}).  The synthetic absorption near 8000~\AA\ is
  produced by O~I \lam8446 but the neighboring observed absorptions
  are produced by the Ca~II IR3.}
\end{figure}

\begin{figure}
\includegraphics[width=.8\textwidth,angle=0]{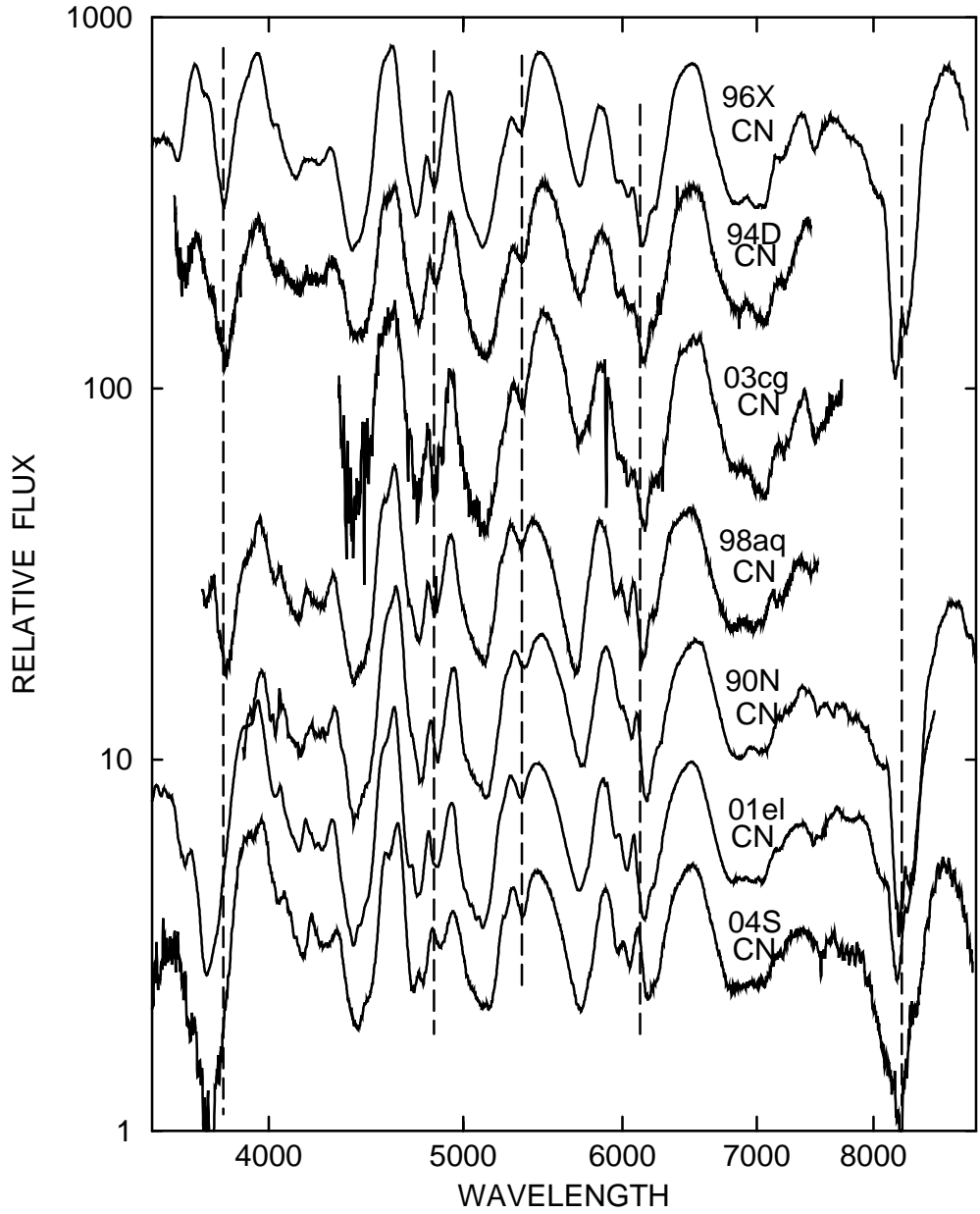}
\caption{Like Fig.~2 but for the seven CNs of the 3 week
postmax sample, with an additional dashed line at 5350~\AA.}
\end{figure}


\begin{figure}
\includegraphics[width=.8\textwidth,angle=270]{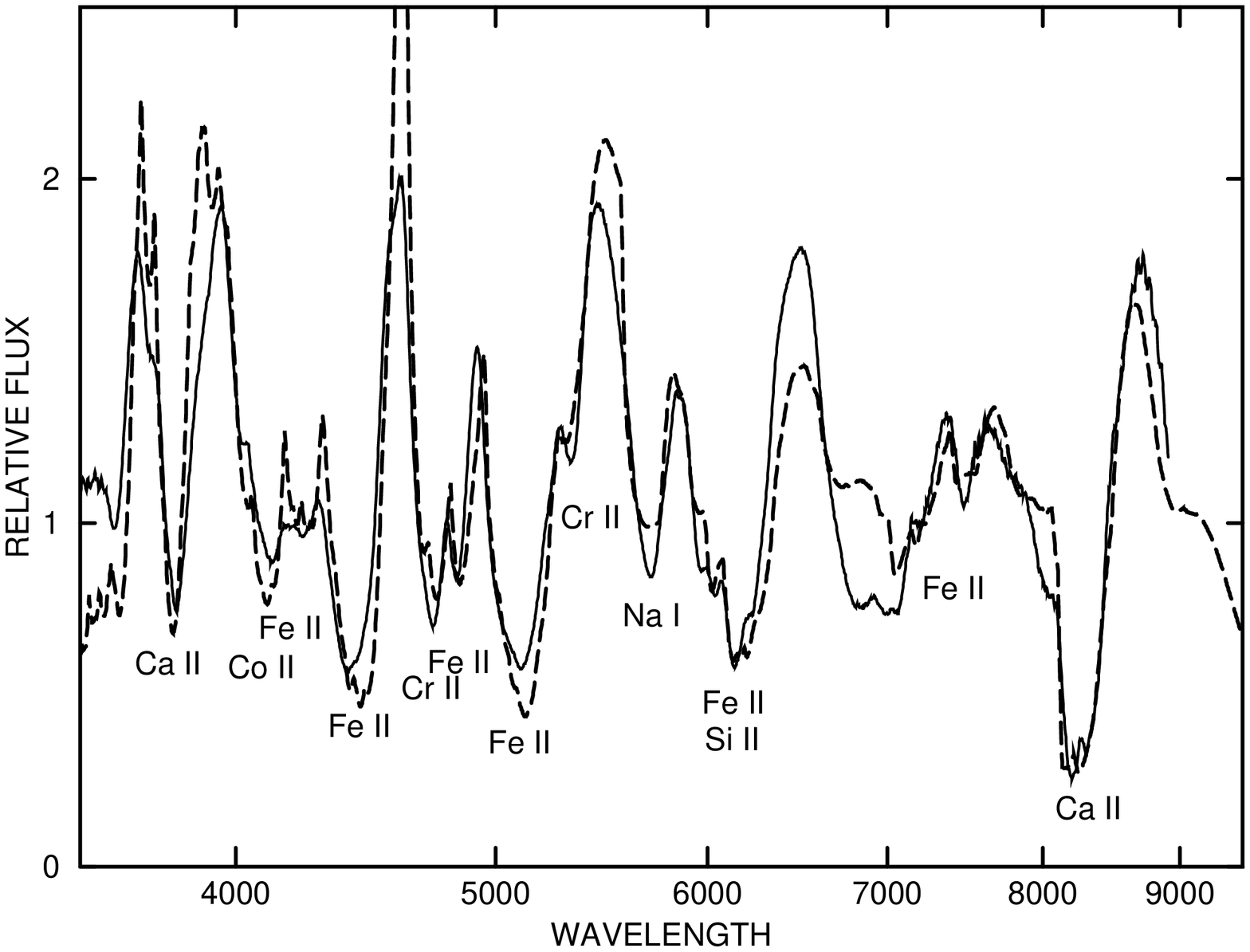}
\caption{The 3 week--postmax spectrum of the CN SN~1996X
  ({\sl solid line}) from Salvo et~al. (2001) compared with a
  synthetic spectrum ({\sl dashed line}).}
\end{figure}

\begin{figure}
\includegraphics[width=.8\textwidth,angle=0]{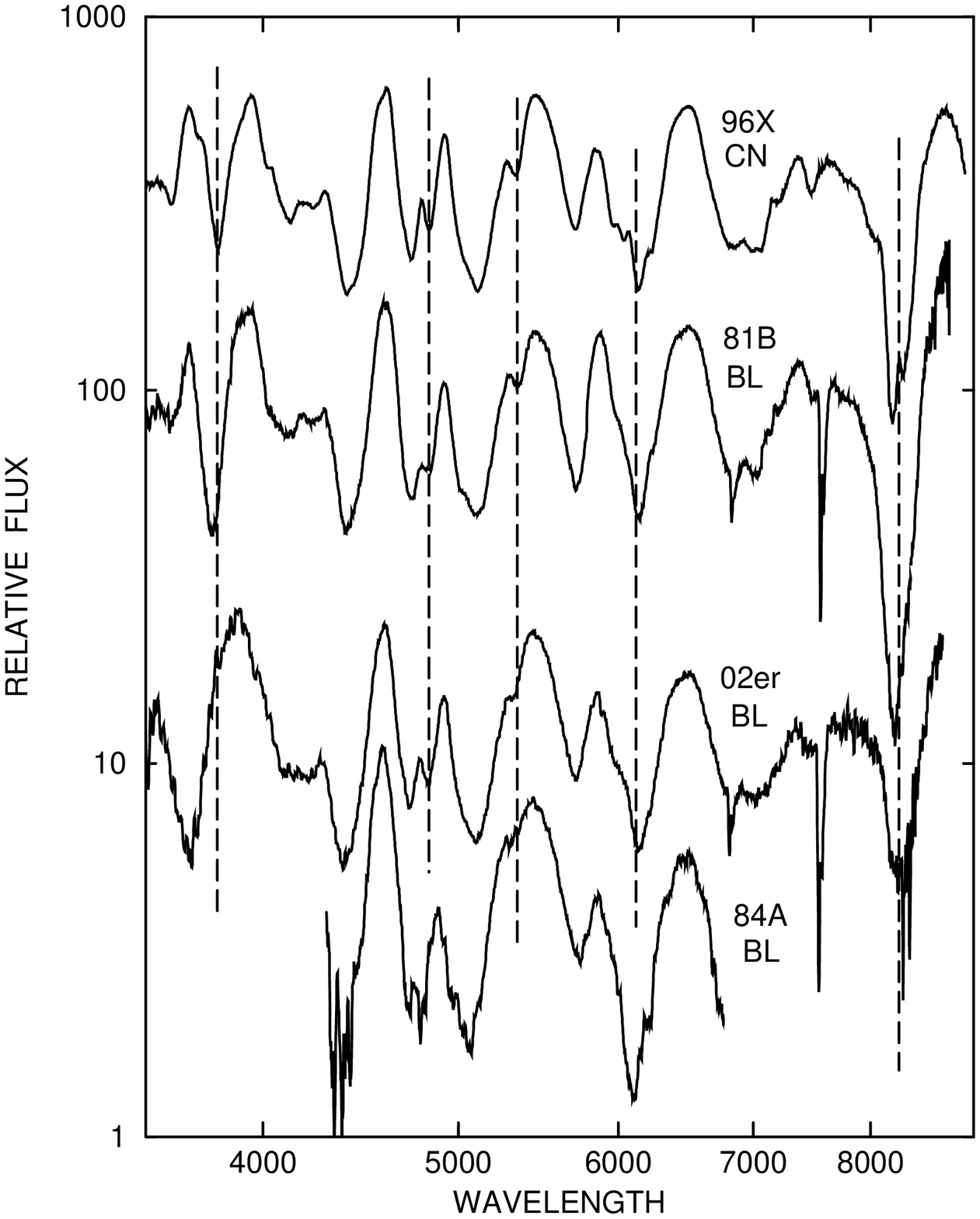}
\caption{Like Fig.~2 but for one CN and the three BLs
of the 3 week postmax sample, with an additional dashed line at
5350~\AA.}
\end{figure}

\clearpage

\begin{figure}
\includegraphics[width=.8\textwidth,angle=0]{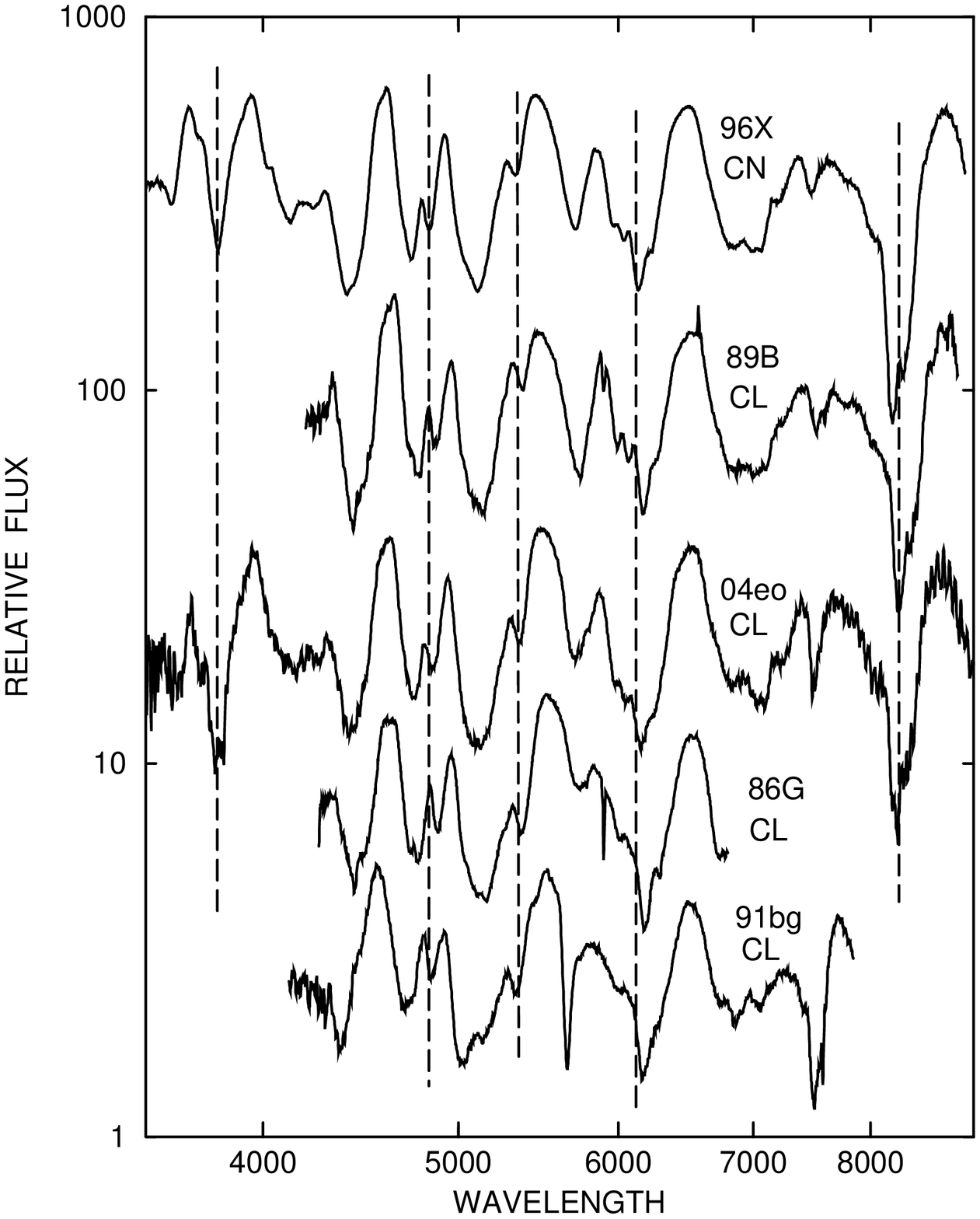}
\caption{Like Fig.~2 but for one CN and the four CLs of the
3 week postmax sample, with an additional dashed line at 5350~\AA.}
\end{figure}


\clearpage

\begin{figure}
\includegraphics[width=.8\textwidth,angle=270]{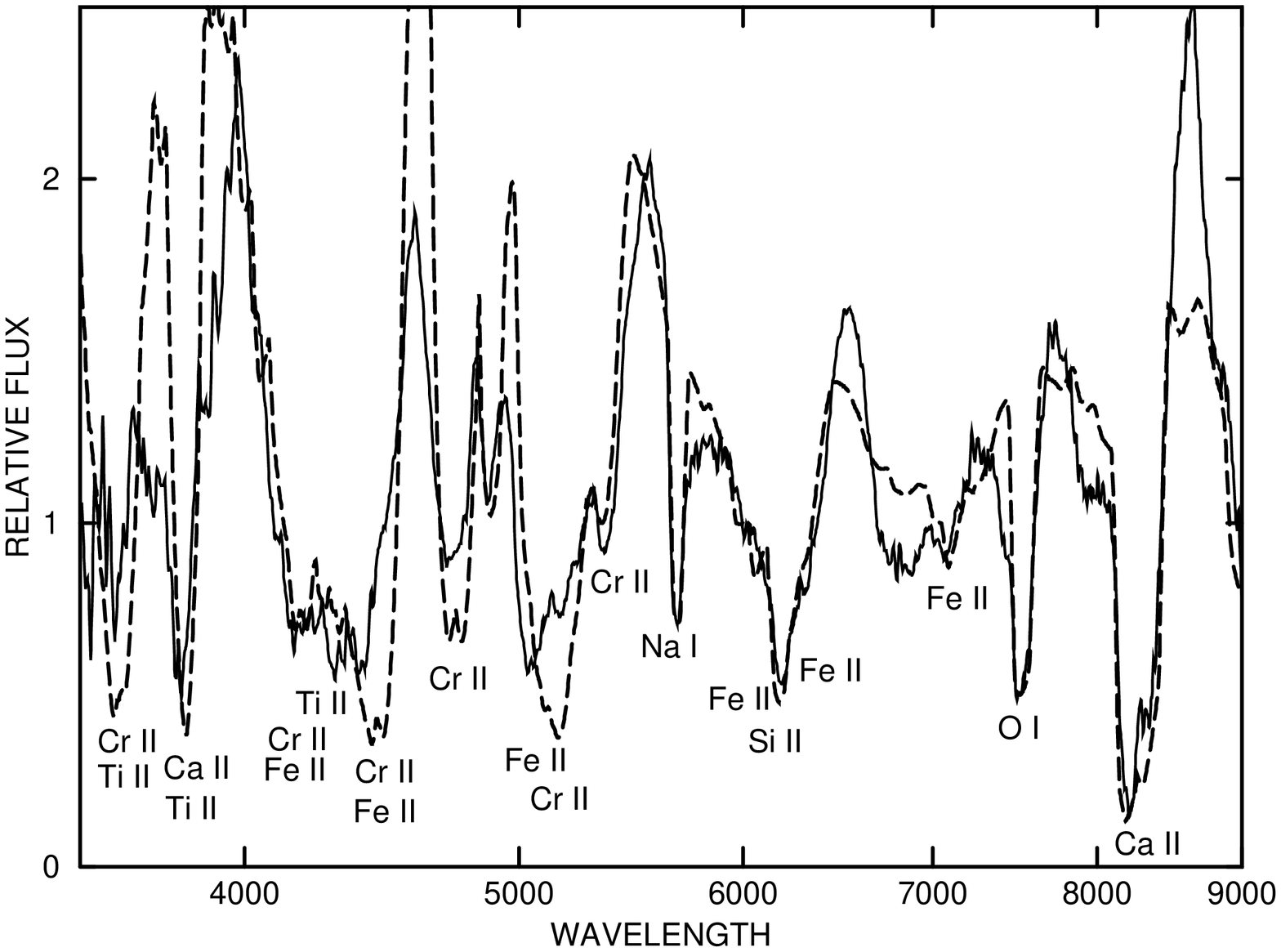}
\caption{The 3 week--postmax spectrum of the CL SN~1991bg({\sl solid
  line}) from Filippenko et~al. (1992a) compared with a synthetic
  spectrum ({\sl dashed line}).}
\end{figure}

\clearpage

\begin{figure}
\includegraphics[width=.8\textwidth,angle=0]{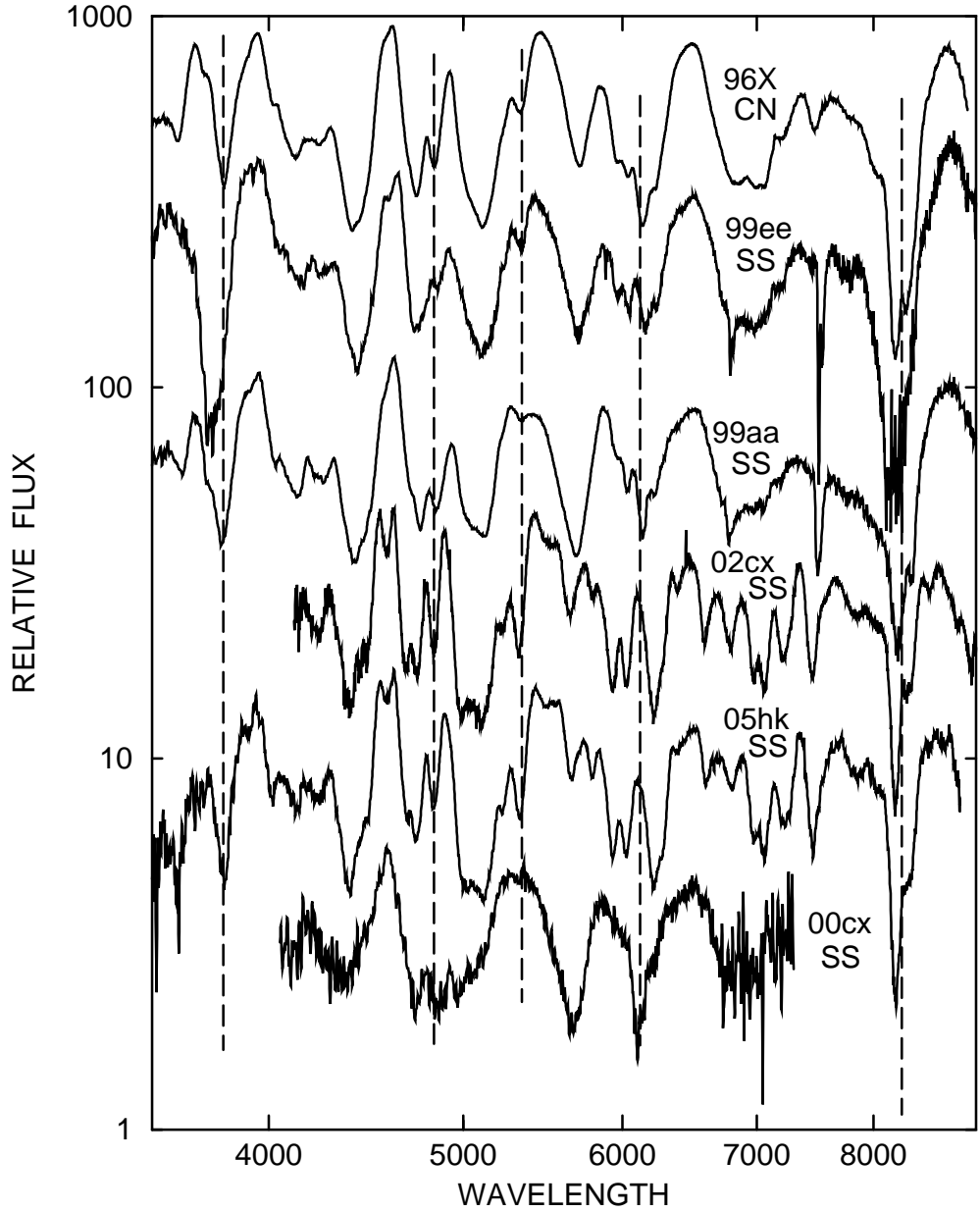}
\caption{Like Fig.~2 but for one CN and the five SSs of the 3 week
postmax sample, with an additional dashed line at 5350~\AA.  The
spectra of SN~2002cx and SN~2005hk have been artificially blueshifted
by 5000 \kms.}
\end{figure}

\clearpage

\begin{figure}
\includegraphics[width=.8\textwidth,angle=270]{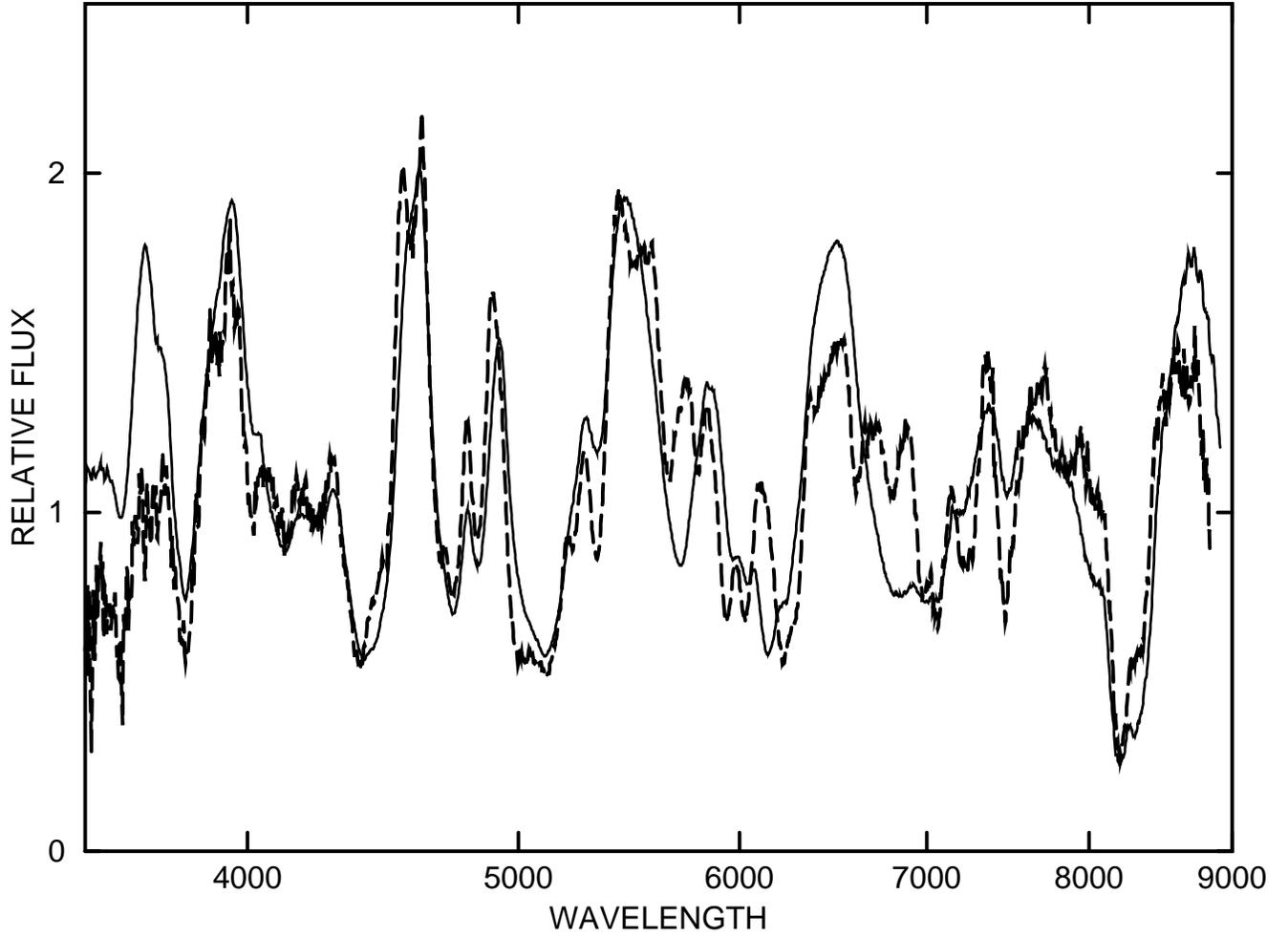}
\caption{ The 3 week--postmax spectra of the SS SN~2005hk ({\sl solid
  line}) from Stanishev et~al. (2007) but artificially blueshifted by
  5000 \kms, and the CN SN~1996X ({\sl dashed line}) from Salvo
  et~al. (2001) are compared.}
\end{figure}

\begin{figure}
\includegraphics[width=.8\textwidth,angle=270]{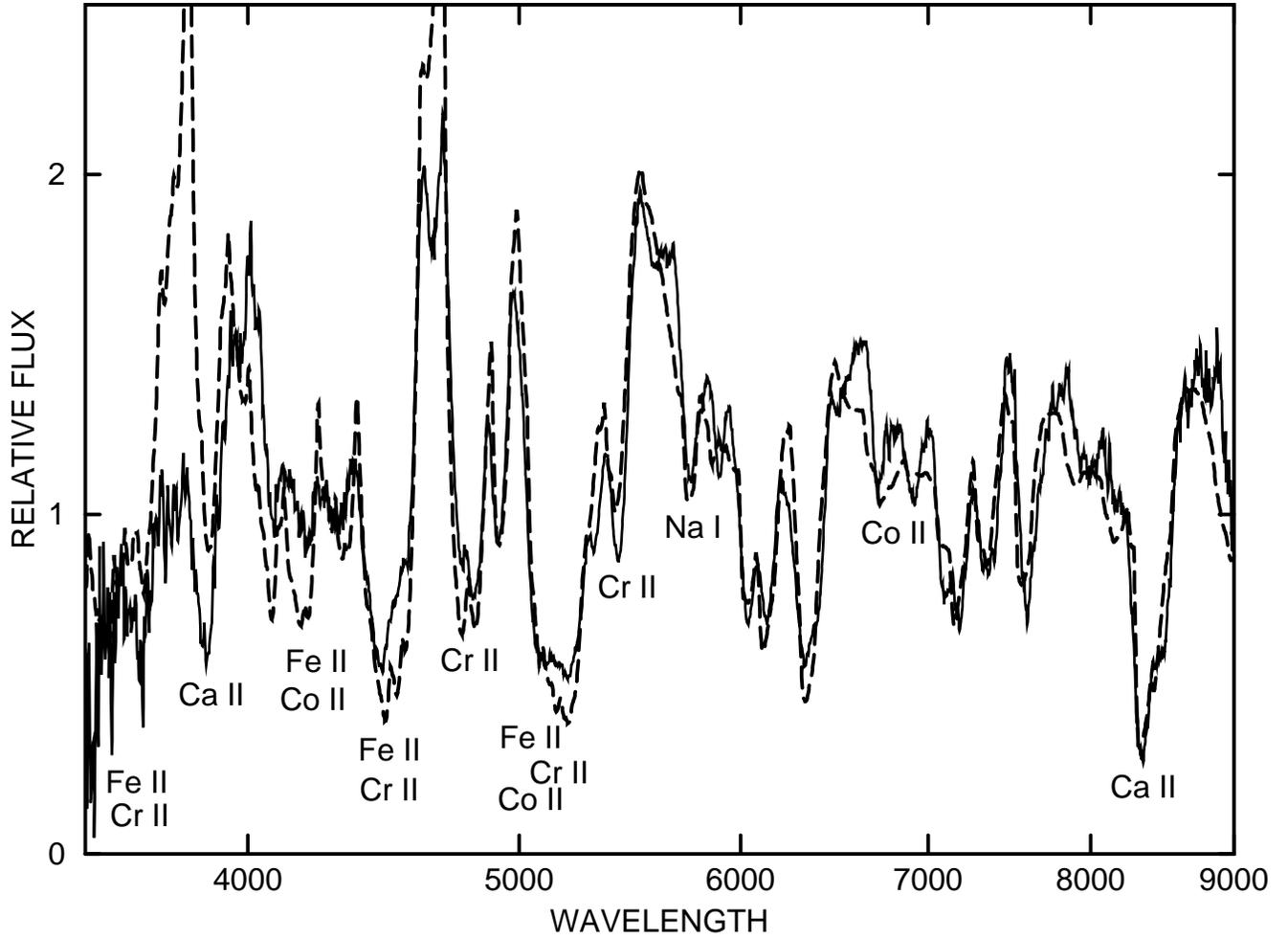}
\caption{The 3 week--postmax spectrum of the SS SN~2005hk ({\sl solid
  line}) from Stanishev et~al. (2007) compared with a synthetic
  spectrum ({\sl dashed line}).  Unlabelled absorption features in the
  synthetic spectrum are produced by Fe~II.}
\end{figure}

\begin{figure}
\includegraphics[width=.8\textwidth,angle=0]{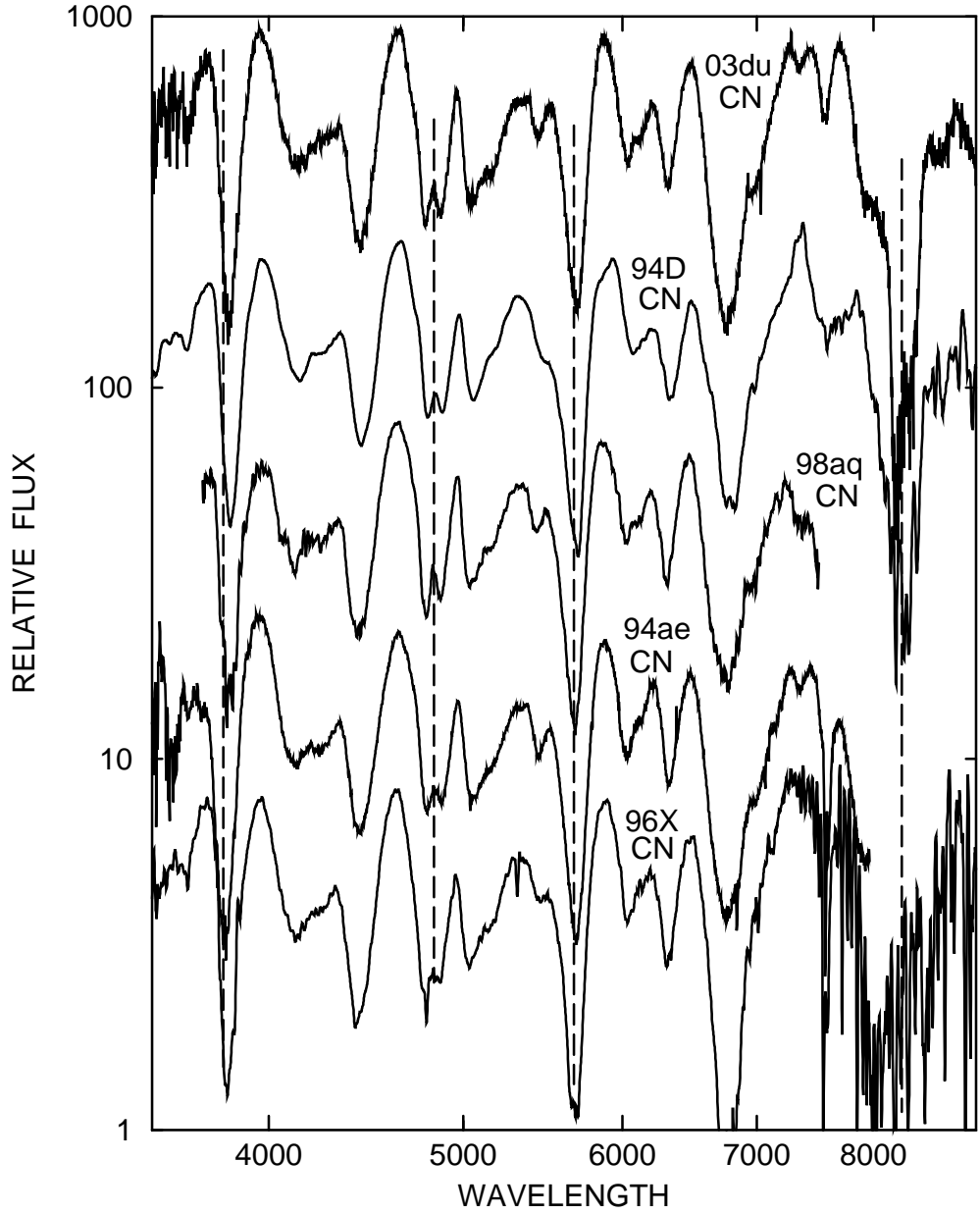}
\caption{Like Fig.~2 but for the five CNs of the 3 month
postmax sample.}
\end{figure}

\begin{figure}
\includegraphics[width=.8\textwidth,angle=270]{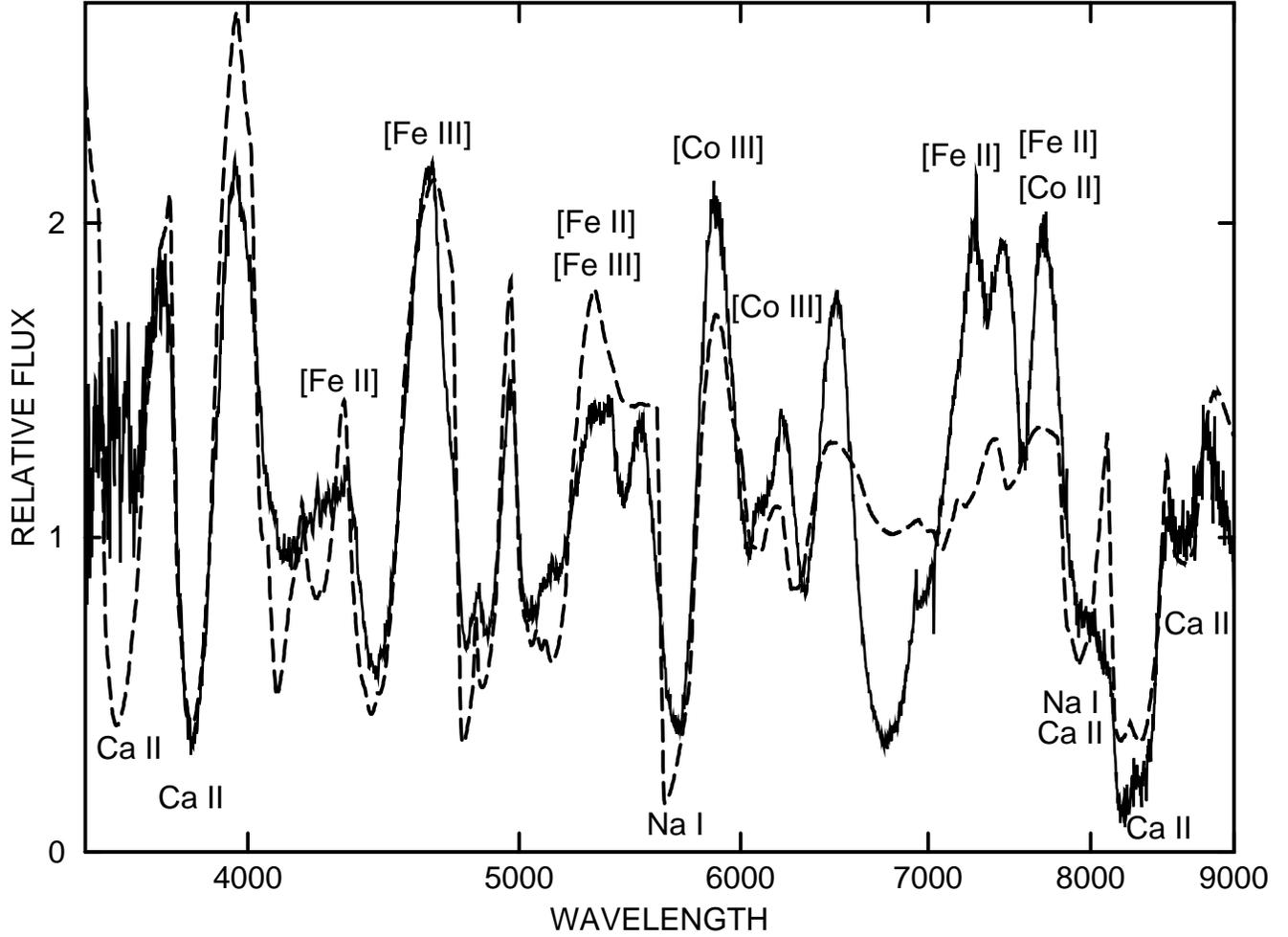}
\caption{The 3 month--postmax spectrum of the CN SN~2003du ({\sl solid
  line}) from Stanishev et~al. (2007) compared with a synthetic
  spectrum ({\sl dashed line}).  The forbidden--line identifications
  that appear above some of the flux peaks are from Bowers
  et~al. (1997) but these are not used in the SYNOW synthetic
  spectrum.  Unlabelled absorption features in the synthetic spectrum
  are produced by Fe~II.}
\end{figure}

\begin{figure}
\includegraphics[width=.8\textwidth,angle=0]{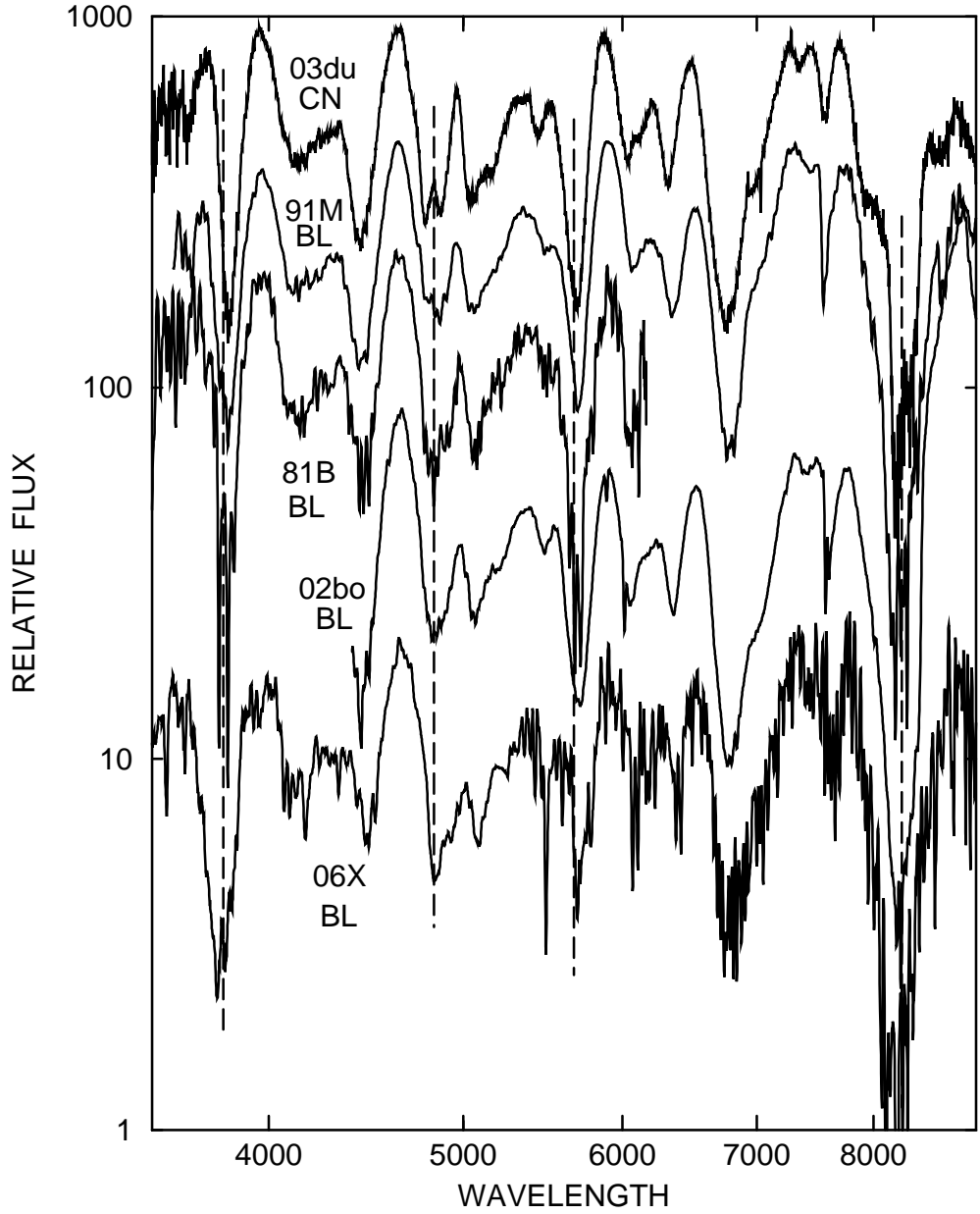}
\caption{Like Fig.~2 but for one CN and the four
BLs of the 3 month postmax sample.}
\end{figure}

\begin{figure}
\includegraphics[width=.8\textwidth,angle=0]{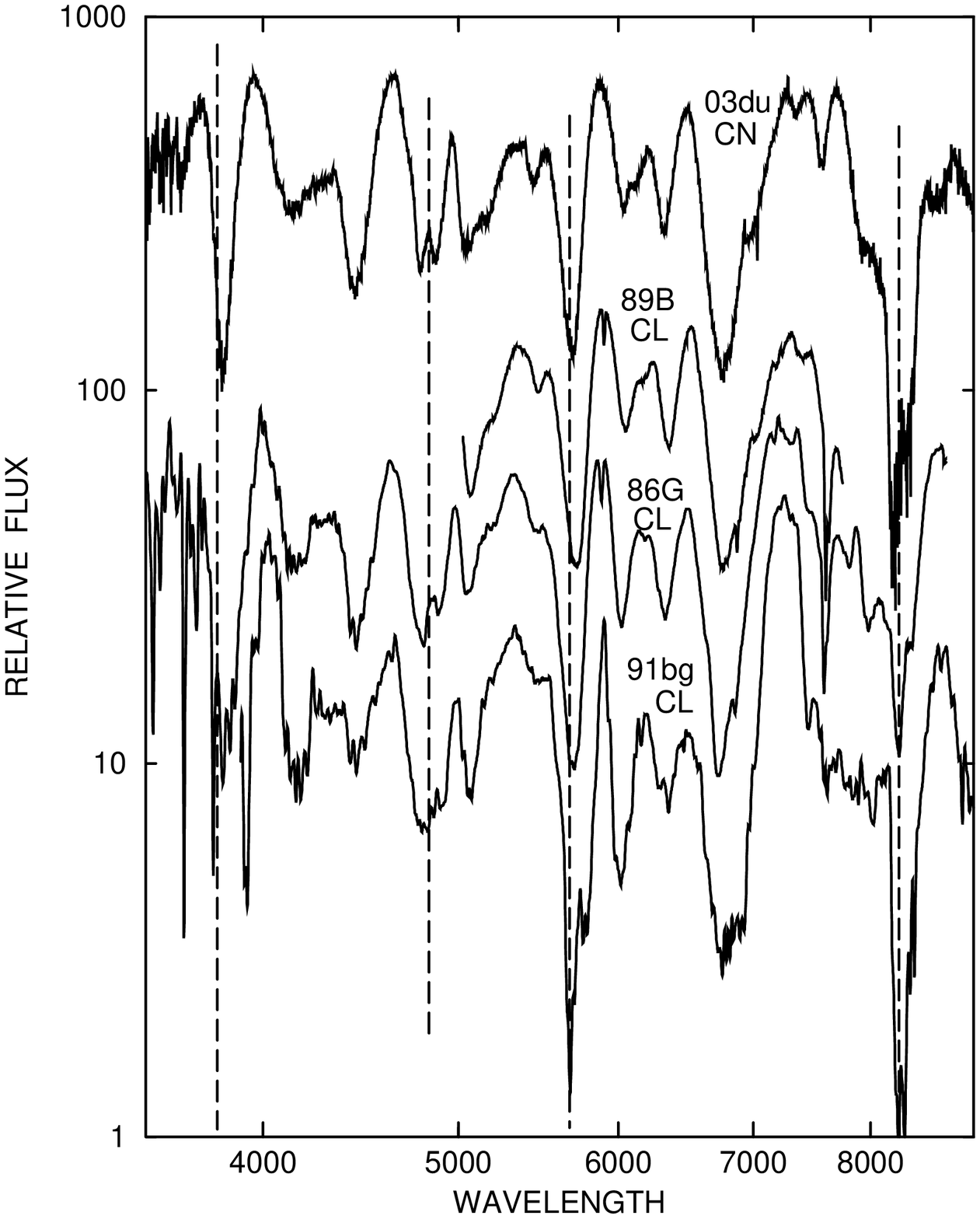}
\caption{Like Fig.~2 but for one CN and the three
CLs of the 3 month postmax sample.}
\end{figure}

\begin{figure}
\includegraphics[width=.8\textwidth,angle=270]{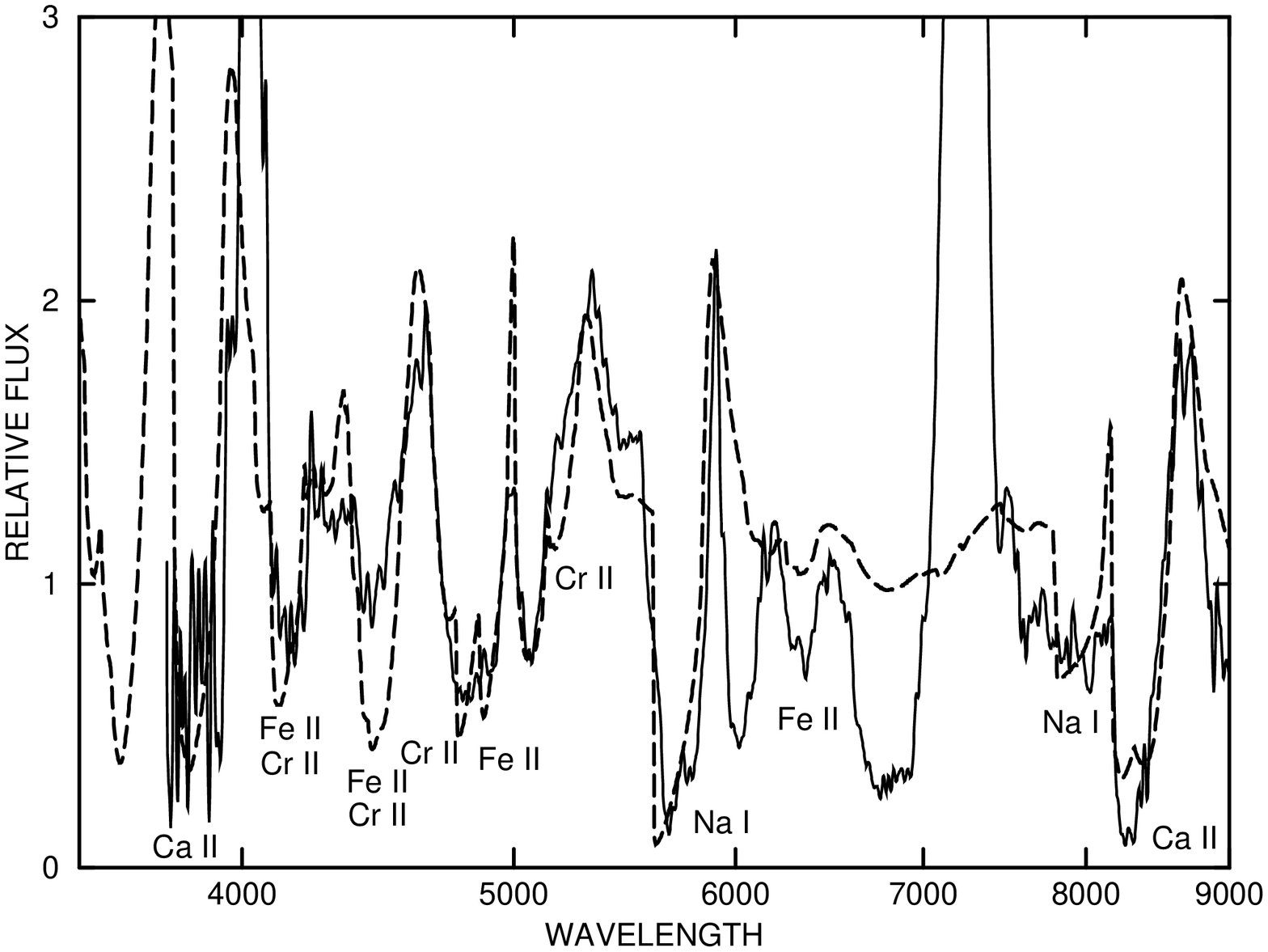}
\caption{The 3 month--postmax spectrum of the CL SN~1991bg ({\sl
  solid line}) from Filippenko et~al. (1992a) compared with a synthetic
  spectrum ({\sl dashed line}).}
\end{figure}

\begin{figure}
\includegraphics[width=.8\textwidth,angle=0]{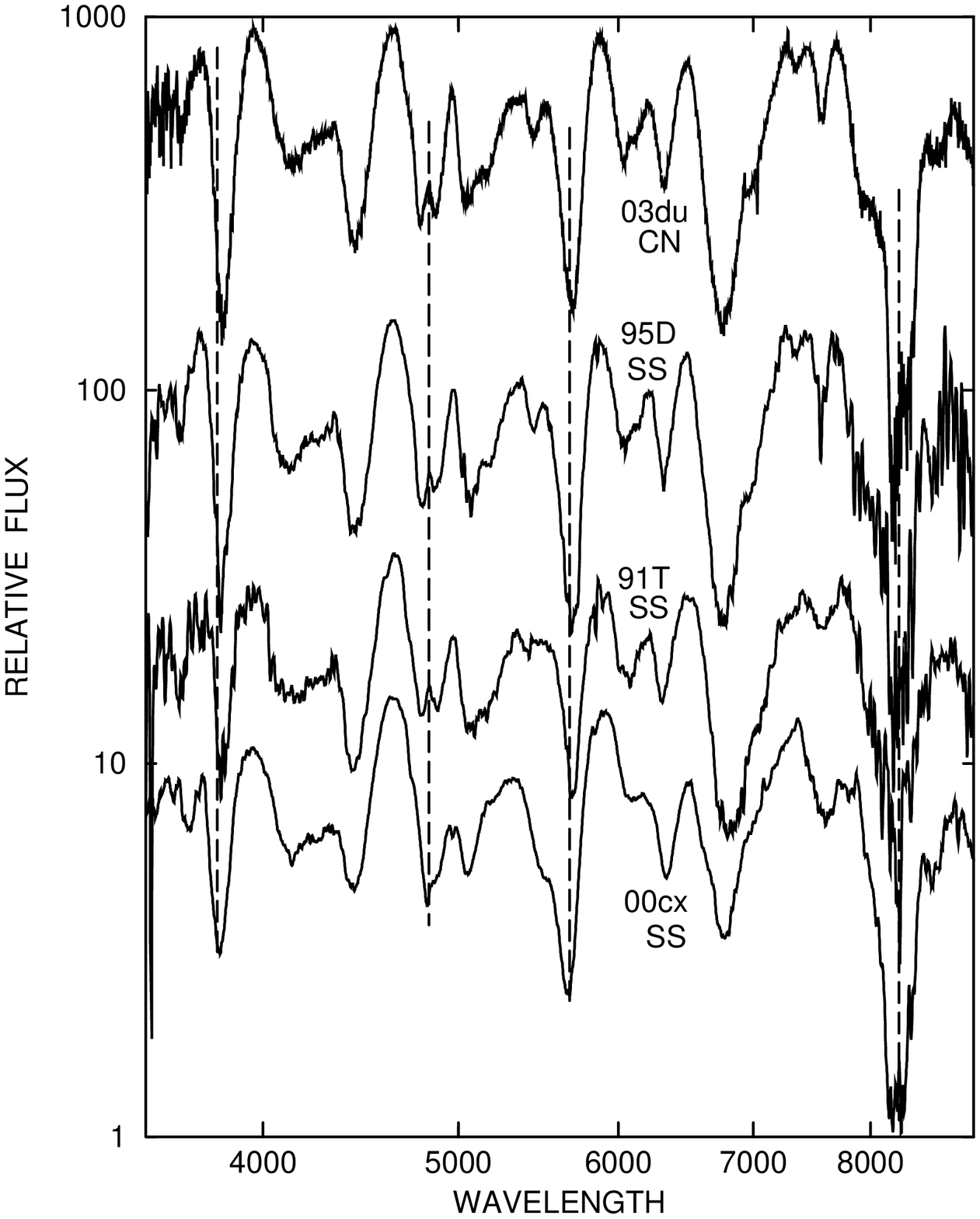}
\caption{Like Fig.~2 but for one CN and the three
SSs of the 3 month postmax sample.}
\end{figure}


\clearpage

\begin{deluxetable}{lcll}
\tablenum{1}

\setlength{\tabcolsep}{4pt}

\tablecaption{The SN~Ia Sample}

\tablehead{\colhead{SN} & \colhead{Epochs} & \colhead{Galaxy} &
\colhead{References} \\

\colhead{} & \colhead{(days)} & \colhead{} & \colhead{} }

\startdata

1981B BL & 22, 93 & NGC 4526 & Branch et al. 1983\\

1984A BL & 8, 19 & NGC 4419 & Barbon et al. 1989\\

1986G CL & 21, 90 & NGC 5128 & Cristiani et al. 1992\\

1989B CL & 8, 19, 92 & NGC 3627 & Wells et al. 1994\\

1990N CN & 7, 21 & NGC 4639 & 7: Leibundgut et al. 1991; 21:
Filippenko et~al. 1992b\\

1991M BL & 81 & IC 1151 & Gomez \& Lopez 1998\\

1991T SS & 83 & NGC 4527 & A. V. Filippenko, unpublished\\

1991bg CL & 19, 91 & NGC 4374 & Filippenko et al. 1992a\\

1992A BL & 6 & NGC 1380 & Kirshner et al. 1993\\

1994D CN & 7, 19, 87 & NGC 4526 & 7, 19: Patat et al. 1996; 87:
Filippenko 1997\\

1994ae CN & 89 & NGC 3370 & Bowers et~al. 1997\\

1995D  SS & 96 & NGC 2962 & Bowers et~al. 1997\\

1996X  CN & 7, 22, 87 & NGC 5061 & Salvo et al. 2001\\

1997br SS & 8 & ESO 576-G40 & Li et~al. 1999\\

1998aq CN & 7, 21, 91 & NGC 3982 & Branch et al. 2003\\

1998bu CN & 8 & NGC 3368 & Jha et al. 1999\\

1999aa SS & 6, 19 & NGC 4469 & Garavini et~al. 2004\\

1999ac SS & 8 & NGC 2848 & Garavini et~al. 2005\\

1999by CL & 7 & NGC 2841 & Garnavich et al. 2004\\

1999ee SS & 22 & IC 5179 & Hamuy et al. 2002\\

2000cx SS & 7, 20, 89 & NGC 524 & Li et~al. 2001\\

2001el CN & 20 & NGC 1448 & Wang et al. 2003\\

2002bf BL & 7 & \nodata & Leonard et al. 2005\\

2002bo BL & 82 & NGC 3190 & Benetti et al. 2004\\

2002cx SS & 21 & \nodata & Li et~al. 2003\\

2002er BL & 6, 20 & UGC 10743 & Kotak et al. 2006\\

2003cg CN & 7, 23 & NGC 3169 & Elias--Rosa et al. 2006\\

2003du CN & 7, 84 & UGC 9391 & Stanishev et al. 2007\\

2004S CN & 8, 19 & \nodata & Krisciunas et~al. 2007\\

2004eo CL & 7, 21 & NGC 6928 & Pastorello et~al. 2007\\

2005hk SS & 21 & UGC 272 & Stanishev et~al. 2007\\

2006X BL & 6, 98 & NGC 4321 & Wang et~al. 2007\\

2006gz SS &7 & IC 1277 & Hicken et~al. 2007\\

\enddata
\end{deluxetable}


\clearpage

\begin{deluxetable}{lccccccc}
\rotate
\tablenum{2}
\setlength{\tabcolsep}{4pt}

\tablecaption{Fitting Parameters for Selected Spectra of the 1 Week
  Postmax Sample}

\tablehead{\colhead{Parameter} & \colhead{SN~1996X} &
\colhead{SN~2004S} &  \colhead{SN~2002bf} &
\colhead{SN~1999by} & \colhead{SN~1999ac} & \colhead{SN~2000cx} \\

\colhead{} & \colhead{CN} & \colhead{CN}  & \colhead{BL} &
\colhead{CL} & \colhead{SS} & \colhead{SS} }

\startdata

\vphot\ (\kms) & 11,000 & 7000 &  11,000 & 6000 & 6000 &
11,000  \\

$\tau$(O~I) & 0.25/[17] & 0.2/[16]   & 0.2/[15] & 0.3/[14] &
0.1/[16] & 0.2/[17] \\

$\tau$(Na~I) & 0.3/[15] & 0.3/[12]  & 0.5/[14] & \nodata & 0.1/[14]
& 0.2/[13] \\

$\tau$(Mg~I) & \nodata & \nodata &  \nodata & 1/[13] & \nodata &
\nodata \\

$\tau$(Mg~II) & 0.6/[15] & 0.2/[13] &  \nodata & \nodata & 0.5/[14] &
\nodata \\

$\tau$(Si~II) & 2/[15] & 0.5/[13] &  3/[21] & 1.2/[12] & 0.5/[14]
& 1/13[15] \\

$\tau$(S~II) & 0.8/[13] & 0.3/[11] &  0.2/[15] & \nodata & 0.3/[10] &
0.3/13[15]\\

$\tau$(Ca~II) & 30/[17] & 15/[24]  & 30/[23] & 100/[14] &
25/[20] & 2/[23] \\

$\tau$(Sc~II) & \nodata &  \nodata & \nodata & 2/[11] & \nodata &
\nodata \\

$\tau$(Ti~II) & \nodata & \nodata & \nodata & 0.8/[11] & \nodata &
0.2/23[25] \\

$\tau$(Cr~II) & \nodata & \nodata & \nodata & 2/[11] & \nodata &
0.3/21[23] \\

$\tau$(Fe~II) & 1/[15] & 0.8/[13]  & 1.5/[21] & 7/[11] & 0.8/[15] &
0.3/19[22] \\

$\tau$(Fe~III) & 0.3/[15] & 0.3/[13]  & \nodata & \nodata & \nodata &
0.3/13[15] \\

$\tau$(Co~II) & 0.6/[15] & 0.6/[13] & \nodata & \nodata & \nodata
& \nodata \\

$\tau$(Ni~II) & \nodata & \nodata & \nodata & \nodata & \nodata &
0.1/13[15] \\

\enddata

\tablenotetext{a}{NOTE.---For each ion, the optical depth $\tau$ is
the optical depth at the photosphere or detachment velocity of the
ion's reference line (ordinarily the ion's strongest line in the
optical spectrum).  Minimum and maximum velocities (in units of
1000~\kms) are preceded by a forward slash, with maximum velocities in
square brackets.}

\end{deluxetable}

\begin{deluxetable}{lcccc}
\rotate
\tablenum{3}
\setlength{\tabcolsep}{4pt}

\tablecaption{Fitting Parameters for Selected Spectra of the
  3 Weeks Postmax Sample}

\tablehead{\colhead{Parameter} & \colhead{SN~1996X} &
\colhead{SN~1991bg} & \colhead{SN~2005hk} & \colhead{SN~2000cx} \\

\colhead{} & \colhead{CN} & \colhead{CL} & \colhead{SS} & \colhead{SS} }

\startdata

\vphot\ (\kms) & 6000 & 6000 & 4000 &7000\\

$\tau$(O~I) & \nodata & 0.6/8[13] & \nodata &  \nodata   \\

$\tau$(Na~I) & 0.4/7[17] & 0.6/10[13] & 0.2/6[9] & 0.6/[18] \\

$\tau$(Si~II) & 0.7/8[13] & 1.2/7[11] & \nodata & 0.7/9[16] \\

$\tau$(Ca~II) & 50/[16] & 500/9[14] & 50/[9]
& \nodata \\

$\tau$(Ti~II) & \nodata & 3/[11] &  \nodata  &  \nodata \\

$\tau$(Cr~II) & 6/[13] & 10/[11] & 10/[8] & 0.3/[18] \\

$\tau$(Fe~II) & 12/[13] & 10/[11] & 30/[8] & 0.8/[18] \\

$\tau$(Fe~III) & \nodata & \nodata &  \nodata  &  0.5/[15] \\
 
$\tau$(Co~II) & 6/[13] & \nodata &  10/[8] &  \nodata \\

\enddata

\tablenotetext{a}{NOTE.---For column descriptions, see the note to Table~2.}

\end{deluxetable}


\begin{deluxetable}{lcc}
\rotate
\tablenum{4}
\setlength{\tabcolsep}{4pt}

\tablecaption{Fitting Parameters for Selected Spectra of the 3 Months
  Postmax Sample}

\tablehead{\colhead{Parameter} & \colhead{SN~2003du} &
\colhead{SN~1991bg}  \\

\colhead{}& \colhead{CN} & \colhead{CL}  }

\startdata

\vphot\ (\kms) & 7000 & 3000\\

$\tau$(Na~I) & 4/[15] & 4/[15]\\

$\tau$(Ca~II) & 10,000/[14] & 1000/[12] \\

$\tau$(Fe~II) & 8/[12] & 2/[10] \\

$\tau$(Cr~II) & \nodata & 0.5/[10] \\

\enddata
\tablenotetext{a}{NOTE.---For column descriptions, see the note to Table~2.}
\end{deluxetable}

\end{document}